\documentclass[lettersize, journal]{IEEEtran}
\usepackage{amsmath, amsfonts}
\usepackage{amssymb}
\usepackage[ruled, linesnumbered]{algorithm2e}
\usepackage{array}
\usepackage[caption=false, font=footnotesize, labelfont=sf, textfont=sf]{subfig}
\usepackage{textcomp}
\usepackage{stfloats}
\usepackage{url}
\usepackage{verbatim}
\usepackage{graphicx}
\usepackage{cite}
\usepackage{multirow}
\usepackage{makecell}
\usepackage{enumitem}
\usepackage{float}

\begin{document}

\title{Obstacle-Aware Length-Matching Routing \\ for Any-Direction Traces in Printed Circuit Board}

\author{Weijie~Fang,
        Longkun~Guo*,
        Jiawei~Lin,
        Silu~Xiong,
        Huan~He,
        Jiacen~Xu,
        and~Jianli~Chen,
\thanks{Weijie~Fang, Longkun~Guo, and Jiawei~Lin are with the School of Mathematics and Statistics, Fuzhou University, Fuzhou, Fujian, China.}
\thanks{Silu~Xiong and Huan~He are with Hangzhou Huawei Enterprises Telecommunication Technologies Co., Ltd, Hangzhou, Zhejiang, China.}
\thanks{Jiacen~Xu is with Shanghai LEDA Technology Co., Ltd, Shanghai, China.}
\thanks{Jianli~Chen is with the School of Microelectronics, Fudan University, Shanghai, China.}
\thanks{A preliminary version of this paper has been accepted by DAC 2024. The corresponding author is Longkun Guo (lkguo@fzu.edu.cn).}
}



\maketitle

\begin{abstract}
Emerging applications in Printed Circuit Board (PCB) routing impose new challenges on automatic length matching, including adaptability for any-direction traces with their original routing preserved for interactiveness. The challenges can be addressed through two orthogonal stages: assign non-overlapping routing regions to each trace and meander the traces within their regions to reach the target length. In this paper, mainly focusing on the meandering stage, we propose an obstacle-aware detailed routing approach to optimize the utilization of available space and achieve length matching while maintaining the original routing of traces. Furthermore, our approach incorporating the proposed Multi-Scale Dynamic Time Warping (MSDTW) method can also handle differential pairs against common decoupled problems. Experimental results demonstrate that our approach has effective length-matching routing ability and compares favorably to previous approaches under more complicated constraints.
\end{abstract}


\section{Introduction}

\begin{figure}[t]
    \centering
    \includegraphics[width=3.2in]{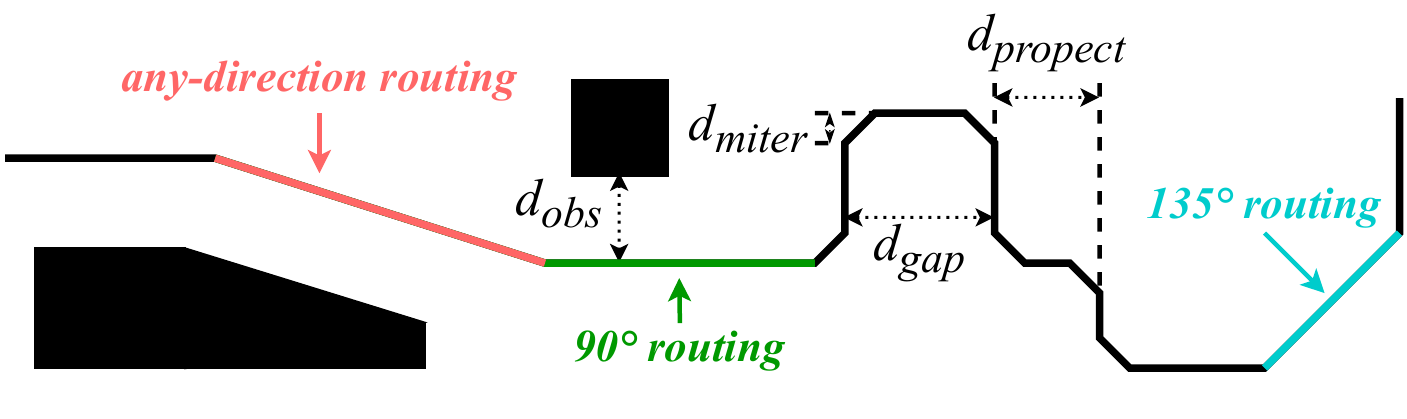}
    \caption{Illustration of routing in various directions and the primary distances restricted in DRC. Solid polygons in the figure denote obstacles.}
    \label{drc}
\end{figure}

\IEEEPARstart{S}{everal} protocols in Printed Circuit Board (PCB) designs demand some parallel signals in the same group to arrive at their destination simultaneously. Mismatching the arrival timing of these signals may result in a critical clock skew that harms the stability and functionality of designs, which derives length-matching techniques to minimize the difference between their propagation delay by matching the length of their traces. Many current length-matching tools still require manual assistance to resolve detail routing in obstacle-dense regions. Meanwhile, length-matching approaches for high-speed designs need to fit routing in any direction, including but not limited to the traditional routing in 90$^\circ$ or 135$^\circ$, which is illustrated in Fig.~\ref{drc}.

\subsection{Related Works}

Much research has been conducted on length matching in recent years, while various approaches have proven their effectiveness in tackling different specific difficulties.

In the aspect of obstacle awareness, many existing methods use a gridded strategy to previously fix safety tracks that do not intersect with obstacles and then determine the detailed routing of traces on these tracks. 
Kohira et al. employed a gridded approach based on the biconnected component to evaluate the upper bound of meandering in space with obstacles and proposed a routing algorithm to approximate this upper bound \cite{kohira2009fast}. Then, they further proposed a heuristic Connectivity Aware Frontier Exploration (CAFE) router that can achieve a small error of length matching in most cases \cite{kohira2010cafe}. 
Yan et al. \cite{Yan2011Obstacle} applied an obstacle-aware region division method on gridded space and used a shortest routing path generator for length matching. 
Hsu et al. \cite{Hsu2019dag} and Chen et al. \cite{Chen2019March} presented several approaches for the clustering, connection, and rip-up and rerouting of bus routers under the obstacle environment. 
Cheng et al. \cite{Cheng2020Obstacle} optimized routability and trace length in their obstacle-avoiding bus length matching approach. 
Yan et al. \cite{Yan2022Single} presented a single-layer obstacle-aware bus router that can minimize the used gridded space and satisfy the length-matching constraints.

\IEEEpubidadjcol

For the determination of a better length matching target.
Kubo et al. \cite{Kubo2004Equidistance} approached length matching with a symmetric slant grid interconnect scheme to prevent a too-large target length.
Nakatani et al. \cite{nakatani2015length} was dedicated to optimizing raw trace before length matching by employing a minimum cost maximum flow algorithm to reduce the maximum trace length while keeping the minimum total trace length.
Ozdal et al. presented an extra resource distribution scheme during the original routing stage to support the possible following length-matching \cite{Ozdal2006Algorithmic}. Also, they introduced Lagrangian Relaxation into length matching that attempts to allocate grid cells to traces with different priorities and minimize the target length \cite{Ozdal2006length}.
Kito et al. \cite{Kito2018fast} introduced simulated annealing into length matching to minimize trace length increment during meandering.
Zhang et al. \cite{Zhang2013parallel} used virtual boundary to fix pins and divided and processed routing space separately, resulting in the reduction of total trace length and a higher similarity among trace routing compared with a previous method \cite{Tsai2011routing}.

Yan et al. \cite{Yan2009Bsg} introduced Bounded-Sliceline Grid (BSG) \cite{Nakatake1998Module} that converts length matching into a quadratic programming problem, followed by a pattern generating rule to achieve the final meandering.
Tseng et al. \cite{Tseng2015Ilp} chose Integer Linear Programming (ILP) to solve length-matching problems, which set the gap between patterns as large as possible to reduce the influence of crosstalk.
Based on the maximum common subsequence of the disordered pin-pairs, Zhang et al. \cite{Zhang2015length} adopted a single commodity flow algorithm \cite{MEDHI2018114} considering the shortest path to resolve the original routing and employed R-flip and C-flip \cite{kohira2009fast} to adjust trace length.
Sato et al. \cite{sato2020fast} presented a pattern generator for set-pair routing by selecting and connecting pin-pairs that improve the efficiency of length matching.
Lee et al. \cite{Lee2013simultaneous} studied a mature simultaneous escape routing algorithm \cite{Luo2010Bescape} and combined it with the length matching of differential pairs based on min-cost median points \cite{Tai2012Escape}.

\subsection{Motivations}

Most existing works on automatic length matching may override the original routing of traces or assume traces are routed in up to eight directions. However, many high-speed PCBs nowadays are designed with traces routed in any direction, and it is usually specified to route such any-direction traces. Leading industrial commercial tools, like Allegro PCB Designer \cite{APD}, specially implement a route offset function to generate such traces. 

In applications, the routing of these traces is hoped to be preserved after length matching because users do not prefer a result that confuses their recognition and interaction, or seriously corrupts their previous specific routing during the Computer-Aided Design (CAD) process. Besides, a trace usually passes different Design Rule Areas (DRA), demanding the length matching approaches to consider multiple Design Rules Checking (DRC). These gaps motivate length-matching techniques to keep up with the industrial standard.

\subsection{Contributions}

\begin{figure}[t]
    \centering
    \includegraphics[width=2.9in]{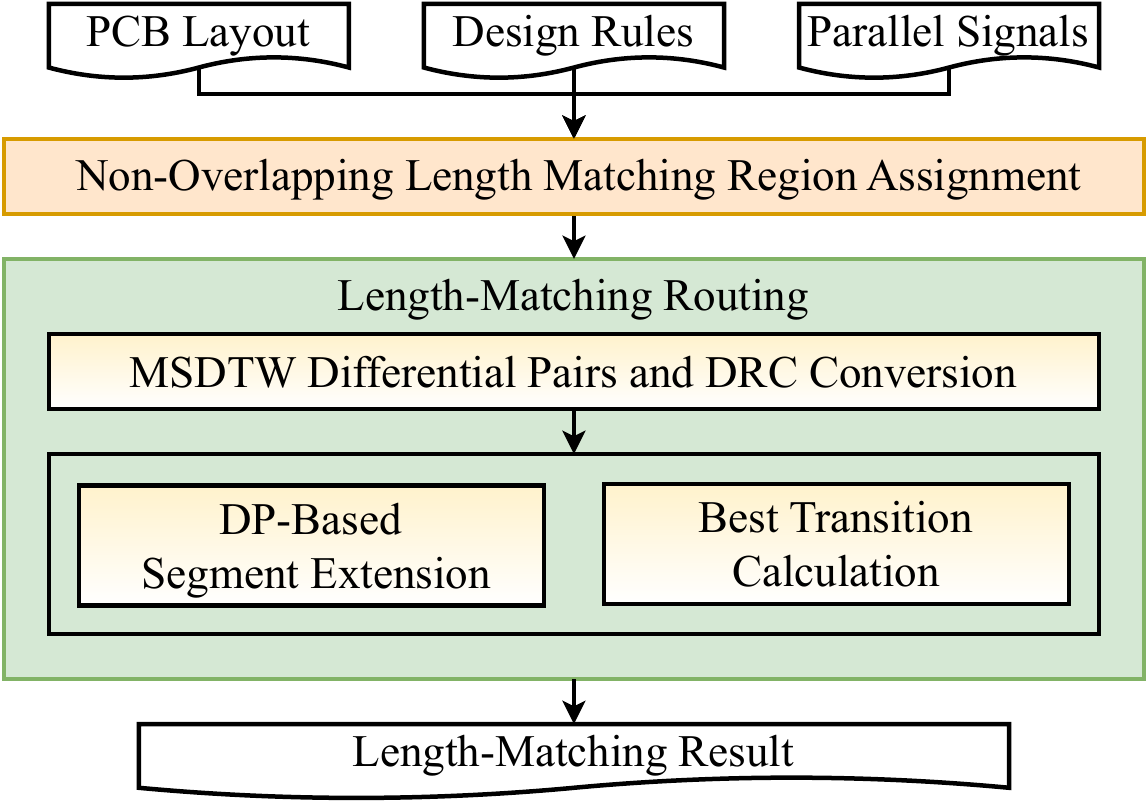}
    \caption{Overview of our length-matching approach.}
    \label{diagram}
\end{figure}

The length-matching process can be divided into two orthogonal stages: assigning non-overlapping regions for original traces and meandering each trace within its own region. This paper mainly focuses on the second stage to achieve automatic length matching while preserving their original properties as much as possible. Meanwhile, applications of length matching frequently involve differential pairs. A differential pair is commonly regarded as a wide single-ended trace during length matching, but this scheme meets many difficulties in practice, especially when the differential pair is not strictly coupled. This paper proposed the Multi-Scale Dynamic Time Warping (MSDTW) method to help tackle these difficulties. Fig.~\ref{diagram} illustrates the algorithmic flow of our approach.

Our contributions are summarized as follows:
\begin{itemize}[leftmargin=10pt]
    \item To the best of our knowledge, this paper is the first length-matching work with respect to arbitrary routing directions, and it supports obstacle-aware routing and multiple DRCs.
    \item The presented length-matching method combines greedy, Dynamic Programming (DP), and computational geometry. Compared with existing approaches, \textit{it routes more flexibly without following fixed tracks or pre-defined modes relying on space regularity}, achieving length matching of any-direction traces concerning original routing.
    \item We proposed the MSDTW method to facilitate the length matching among differential pairs. \textit{It can convert a differential pair into a median trace against several issues in coupling}, and the median trace after length matching can be simply restored to the differential pair. The applications of MSDTW are not limited to our length-matching method.
\end{itemize}

The remainder of this paper is organized as follows.
Section 2 introduces the preliminaries of our works.
Section 3 briefly discusses our region assignment.
Section 4 presents our detailed meandering and MSDTW method.
Section 5 demonstrates the experimental results.
Section 6 concludes this paper.

\section{Problem Formulation}

Length matching is also known as delay tuning because it generally works on the traces already routed in a PCB, and the length of a trace stands for the propagation delay of the signals on it. Although propagation delay is not the only factor to be considered in timing engines, the other delays are generally ignored when discussing length matching. Nevertheless, our approach meanders each trace independently, thereby supporting the individual target lengths of each trace. Rigorously, the influence of other delays can also be considered by adjusting the target length of each trace, i.e., the precise propagation delay that each signal actually needs.

In this paper, we focus on length-matching routing on any-direction traces to meet the emerging industrial requirement from high-speed PCBs nowadays. The clarification of primary distances restricted in DRC about length matching shown in Fig.~\ref{drc} is listed as follows:
\begin{itemize}[leftmargin=39pt]
    \item[$d_{gap}$:] restricts the distance between traces to prevent self-inductance, crosstalk, etc.
    \item[$d_{obs}$:] restricts the distance between a trace and an obstacle.
    \item[$d_{protect}$:] restricts the minimum length to prevent the occurrence of extremely short trace segments.
    \item[$d_{miter}$:] configures the corners mitered for convex patterns. In practice, any rotation of a right angle or an acute angle will be mitered by obtuse angles.
\end{itemize}
Some other important concepts mentioned in this paper are given as follows:
\begin{itemize}[leftmargin=0pt]
    \item[] \textbf{Trace:} trace of a signal consisting of connected segments in PCB layout, also indicated by net or wire.
    \item[] \textbf{Any-direction:} the traces that can be routed not only in 90$^\circ$ or 135$^\circ$ are called any-direction traces.
    \item[] \textbf{Target length $l_{target}$:} a length that a trace in a matching group needs to match, no less than the original length of the trace.
    \item[] \textbf{Routable~area:} the union of non-overlapping routing regions assigned to a trace, represented as some irregular polygons.
    \item[] \textbf{Obstacle:} a polygon that the trace cannot pass, converted into a part of the routable area in this paper.
\end{itemize}
Therefore, the problem we address in this paper is formulated as follows:
\begin{itemize}[leftmargin=0pt]
    \item[] \textbf{Any-direction length-matching problem}: Given a PCB layout, design rules, and matching groups. For each matching group with $l_{target}$, extend each trace in the group utilizing the space of its rouTable~area to make its length equal $l_{target}$, while preserving its original specific routing as much as possible.
\end{itemize}


For digestibility, we use the convex pattern with corners at a right angle in the remaining discussion to omit tedious details of geometry computation.




\section{Region Assignment}

Based on the relation between length and space revealed in \cite{Yan2009Bsg}, we need only assign sufficient regions for each trace to hold feasible length-matching routing. Similar problems have been discussed in many works, such as \cite{Yan2009Bsg} using Quadratic Programming and \cite{Ozdal2006length} using Lagrangian relaxation. For the sake of better fitting our specific requirement, we divide the design according to its layout to compose several regions and consider the following constraints:
\begin{enumerate}[leftmargin=20pt]
    \item Neighbor Validity: A region can only be assigned to its neighbor traces:
    \begin{equation}
         x_{ij} = 0,\ \text{region $i$ is not the neighbor of trace $j$}            
        \label{basic_constraint}
    \end{equation}
    where $x_{ij}$ denotes the space region $i$ assign to trace $j$.
    \item Feasibility: Any assignment of a region space must be positive and bounded by its capacity:
    \begin{equation}
        \sum_{j}{x_{ij}} \leqslant Cap_i,\ x_{ij} \geqslant 0,\ \forall i, j
        \label{feasibility_constraint}
    \end{equation}
    where $Cap_i$ denotes the space capacity of region $i$.
    \item Sufficiency: A trace must receive sufficient space from its neighbor regions:
    \begin{equation}
        \sum_{i}{x_{ij}} \geqslant Req_j,\ \forall i, j
        \label{sufficiency_constraint}
    \end{equation}
    where $Req_j$ denotes the required space for trace $j$.
\end{enumerate}
Here, we employ a Linear Programming (LP) problem to solve this assignment:
\begin{equation}
    \begin{array}{rl}
        \text{\underline{Assignment Problem:}} & \\
                                  \text{find:} & \text{feasible } x_{ij} \\
                            \text{satisfying:} & \text{neighbor validity constraint } \eqref{basic_constraint} \\
                                               & \text{feasibility constraint } \eqref{feasibility_constraint} \\
                                               & \text{sufficiency constraint } \eqref{sufficiency_constraint} \\
    \end{array}
\end{equation}
This assignment scheme ensures the preserved original routing is contained in the rouTable~area for the following stages. Some techniques of existing works can help to figure out a better routing if the LP is infeasible \cite{Chang2019Obstacle}. We are not going to discuss them in detail here.

\section{DP-Based Segment Extension}

In order to increase the length of a trace $l_{trace}$ to its target length $l_{target}$, our routing method inserts convex patterns perpendicular to its segment, this process is called the extension of segments. This extension is held by computational geometry so that it fits any-direction routing. Each segment is extended as much as possible, and a segment after the extension is replaced by several new component segments for further extension if needed. The extension will be conducted iteratively until $l_{trace}$ is within the error tolerance of $l_{target}$, resulting in patterns similar to the combination of Accordion and Trombone. The whole DP-based extension is presented in Alg.~\ref{dp_routing}.

\subsection{State Transition}


Our method optimizes the extension of a segment using a DP algorithm. First of all, we discretize the segment into points using a configurable step length $l_{disc}$. In this paper, we define $d(a, b)$ denotes the Euclidean distance between points $a$ and $b$. Formally, a segment $AB$ with node points $A$ and $B$ will be discretized to a set $U = \{u\ |\ u$ is on segment $AB \}$ of $n$ different points, where $u_1 = A$, $u_n = B$ and $\forall{i \in [2,n-1]}, d(u_{i-1}, u_i) = l_{disc}$. We may slightly increase $d_{gap}$ and $d_{protect}$ or adjust $l_{disc}$ to make the former divisible by the latter.

Based on the discretization above, we define $dp[i][dir]$ to denote the best extension result, in which the patterns are inserted within the previous $i$ points, and the last inserted pattern is in the $dir$ direction of the segment. Without loss of generality, we define the clockwise direction as the positive direction and the counterclockwise direction as the negative direction, represented by ``$1$'' and ``$-1$'', respectively. We set the initial state as:
\begin{equation}
    dp[1][-1] = dp[1][1] = 0
    \label{beginning}
\end{equation}
During the transition, we initialize a new state by:
\begin{equation}
    dp[i][dir] = dp[i - 1][dir],\ i \in [2, n]
    \label{initialization}
\end{equation}
For each pattern with $w$-steps width, whose feet are located in $u_{i-w}$ and $u_i$, respectively, we calculate its maximum valid height $h$ and try to attach it with the best available predecessor states to obtain a better result of the current state. Thereby, we can obtain a simplified state transition equation as follows:
\begin{equation}
    dp[i][dir] = \max(dp[i][dir], dp[i-w][\pm\ dir] + h),\ i \in [2, n] \\
    \label{dp_simplified}
\end{equation}

\begin{algorithm}[t]
    \SetAlgoLined
    \caption{DP-based segment extension.\label{dp_routing}}
    
    \KwIn {$trace$ before length matching; target length $l_{target}$.}
    \KwOut {$trace$ after length matching.}
    maintaining unexpanded segments using queue $Q$ \\
    \While {$l_{trace} \neq l_{target}$ and $|Q| \neq 0$}
    {
        pop segment $seg$ from $Q$, discretize $seg$ into point set $U$ \\
        $dp[1][1] = dp[1][-1] = 0$ \\
        \For {$i \gets 2$ to $n$,\ $dir$ in $\{-1, 1\}$}
        {
            $dp[i][dir] = dp[i-1][dir]$ \\
            \If {$i = n$ \textbf{or} $d(u_i, u_n) \geqslant d_{protect}$}
            {
                \For {$w \gets 1$ to $i$}
                {
                    calculate the maximum $h$ based on width $w$ \\
                    calculate $dp[i][dir]$ considering priority \\
                }
            }
        }
        $dir_{max} = \mathop{\arg\max}\limits_{dir}{dp[n][dir]}$ \\
        \If {$dp[n][dir_{max}] > 0$}
        {
            $l_{trace} = l_{trace} + dp[n][dir_{max}]$ \\
            restore the patterns of the best result \\
            push the new segments replacing $seg$ into $Q$ \\
        }
    }
\end{algorithm}

\begin{figure}[t]
    \centering
    \subfloat[From the same direction.]{\includegraphics[width=1.35in]{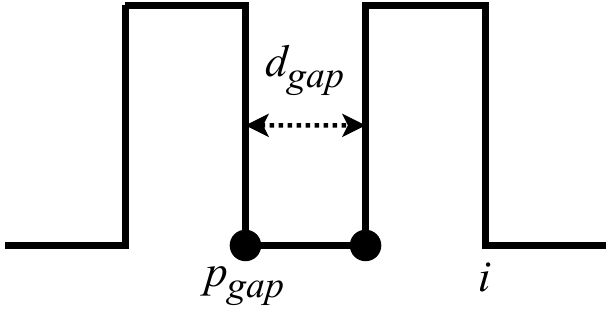}\hspace{0.2in}%
        \label{p_gap}}
    \hfil
    \subfloat[From the opposite direction.]{\hspace{0.2in}\includegraphics[width=1.35in]{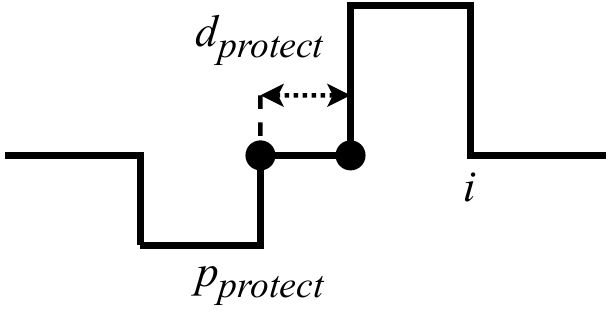}%
        \label{p_protect}}
    \label{gap_protect_fig}
    \centering
    \subfloat[Connect to a pattern.]{\includegraphics[width=1.35in]{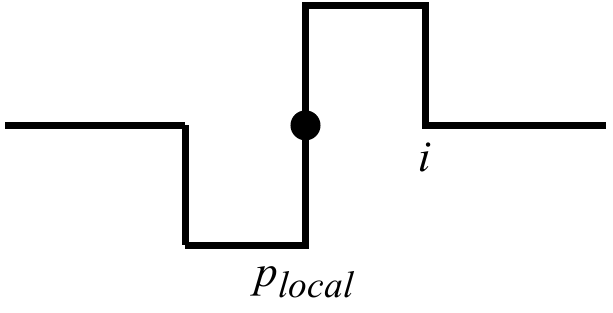}\hspace{0.2in}%
        \label{p_local}}
    \hfil
    \subfloat[Connect to a node point.]{\hspace{0.2in}\includegraphics[width=1.35in]{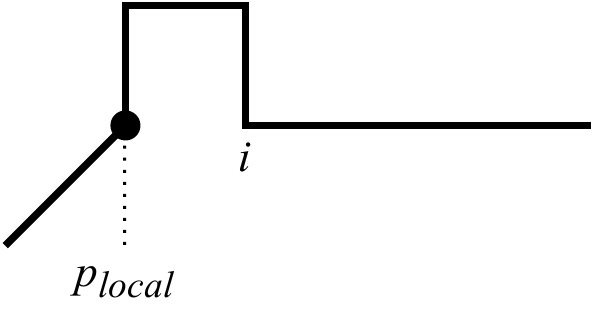}%
        \label{p_node}}
    \caption{Four kinds of valid state transitions.}
    \label{valid_state}
\end{figure}

According to the DRC introduced in Section 2, the actual predecessor states available for the state transition rather than all cases of $dp[i-w][\pm\ dir]$ shall be detailed. If the current state is transited from the one with the same direction, it must keep at least $d_{gap}$ away from the foot of any possibly existing pattern inserted previously. While for the opposite direction, the distance that must be kept is at least $d_{protect}$. Besides, there is a particular valid state transition in which the pattern connects to the previously inserted one or a node of the extended segment. The above four cases are illustrated in Fig.~\ref{valid_state}. Hence, we conclude the actually meaningful states for transition as follows:
\begin{equation}
    \begin{array}{lll}
        dp[i-w][\pm\ dir] = \max\left[
        \begin{array}{lll}
            dp[p_{gap}][dir] \\ 
            dp[p_{protect}][-dir] \\ 
            dp[p_{local}][-dir] \\ 
        \end{array}
        \right],\ i \in [2, n] \\
        \\
        \hspace{-0.05in} \text{where} \\
        \\
        \hspace{0.35in} \left\{
        \begin{array}{lll}
            p_{gap}     & = i - w - \frac{d_{gap}}{l_{disc}} \\
            \vspace{-0.1in} \\
            p_{protect} & = i - w - \frac{d_{protect}}{l_{disc}} \\
            \vspace{-0.05in} \\
            p_{local}   & = i - w,\ \text{need extra condition} \\
        \end{array}
        \right.
    \end{array}
    \label{all_dp_eq}
\end{equation}
Certainly, none of these positions can be less than 1. Otherwise, the corresponding state is invalid.

\begin{figure}[t]
    \vspace{-10pt}
    \centering
    \subfloat[$i$ meets the extra condition of $p_{local}$.]{\hspace{0.1in}\includegraphics[width=1.35in]{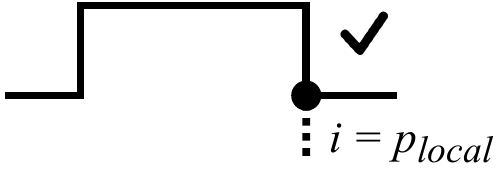}\hspace{0.1in}%
        \label{priority_condition_1}}
    \hfil \hfil
    \subfloat[$i$ does not meet the extra condition of $p_{local}$.]{\hspace{0.1in}\includegraphics[width=1.35in]{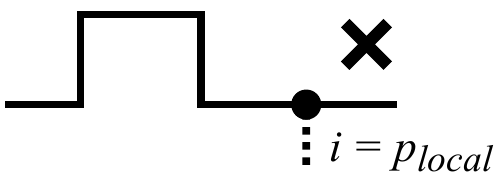}\hspace{0.1in}%
        \label{priority_condition_2}}
    \caption{Illustration of different candidate states with the same value. (a) and (b) contribute the same value to $dp[i][dir]$. Only (a) allows the transition of $p_{local}$, so it has a higher priority than (b) to be retained.
    }
    \label{priority_condition}
\end{figure}
\begin{figure}[t]
    \centering
    \subfloat[Connected.]{\includegraphics[width=1.5in]{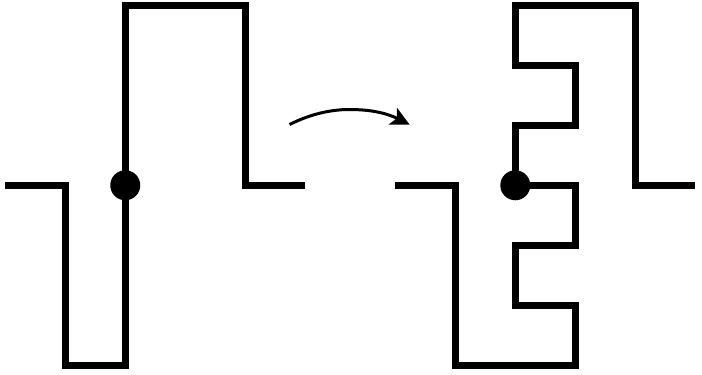}%
        \label{priority_connect_1}}
    \hfil \hfil
    \subfloat[Disconnected.]{\includegraphics[width=1.5in]{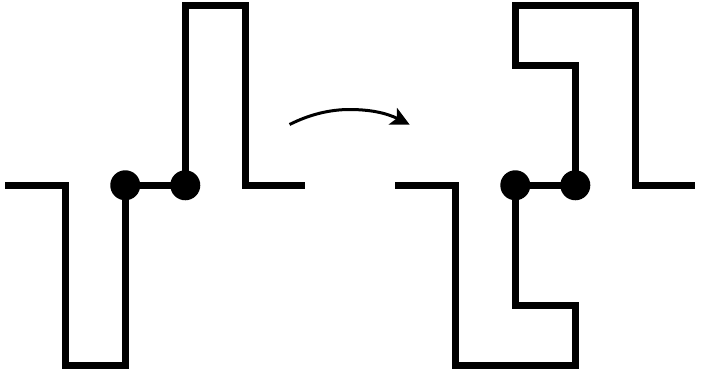}%
        \label{priority_connect_2}}
    \caption{Illustration of why the patterns are hoped to be connected. The original segment has a capacity of only two patterns, so both cases have the same DP result. However, the former case can provide the capacity of an extra pattern.}
    \label{priority_connect}
\end{figure}

If multiple predecessor states have the same value, choosing any of them will not change the final result of $dp[i][dir]$. However, different choices may affect the following states because the transition from $p_{local}$ needs an extra condition illustrated in Fig.~\ref{priority_condition}. Besides, without compromising the current result, the state in which two patterns are connecting will likely bring capacity for extra patterns, as illustrated in Fig.~\ref{priority_connect}, which benefits further meandering on the meandered patterns in possible subsequent iterations. We retain these states as a higher priority during state transitions.


There may come a consideration: To ensure the integrity of state transition, is it necessary to maintain both sub-states of $dp[i][dir]$ for cases without and with a new pattern whose one foot is inserted in the $i$th point? Let us assume $dp[i][dir][0]$ and $dp[i][dir][1]$ denote the above two cases respectively without loss of generality, then Eq.~\eqref{dp_simplified} will be converted to:
\begin{equation}
    \begin{array}{lll}
        \begin{array}{l}
            \left\{
            \begin{array}{l}
                dp[i][dir][0] = \max\left[
                \begin{array}{lllll}
                    dp[i-1][dir][0] \\
                    dp[i-1][dir][1] \\
                \end{array}
                \right] \\
                \\
                dp[i][dir][1] = \max\left[
                \begin{array}{lllll}
                    dp[i-w][\pm\ dir][0] + h \\
                    dp[i-w][\pm\ dir][1] + h \\
                \end{array}
                \right] \\
            \end{array}
            \right. , \\
            \\
            \multicolumn{1}{r}{i \in [2, n]} \\
        \end{array} \\
        \\
        \hspace{-0.05in} \text{where} \\
        \\
        \left\{
        \begin{array}{l}
            dp[i-w][\pm\ dir][0] = \max\left[
            \begin{array}{lllll}
                dp[p_{gap}][dir][0] \\
                dp[p_{protect}][-dir][0] \\
            \end{array}
            \right] \\
            \\
            dp[i-w][\pm\ dir][1] = \max\left[
            \begin{array}{lllll}
                dp[p_{gap}][dir][1] \\
                dp[p_{protect}][-dir][1] \\
                dp[p_{local}][-dir][1] \\
            \end{array}
            \right] \\
        \end{array}
        \right. \\
    \end{array}
    \label{dp_two_cases}
\end{equation}
Observing, except the state $dp[p_{local}][-dir][0]$ is never used because of the invalid transition, the other cases of $dp[i][dir][0]$ and $dp[i][dir][1]$ can be merged into $dp[i][dir]$ since we always choose the maximum one of them before adding an identical $h$. And this merge can be implemented automatically by DP itself if we do not adopt the last dimension to maintain both of the sub-states. For handling the exception of $p_{local}$, we mark each state for whether it is transited through a newly inserted pattern, and this mark is also useful in restoring the final result.

\subsection{Maximum Transition Gain of DP}

To determine the maximum transition gain of the proposed DP, we need to calculate the maximum valid height $h$ of pattern $C$ built on $u_{i-w}$, $u_i$. Notably, that $h$ is valid does not guarantee any height $h' \le h$ if the pattern routes around obstacles because a shrunk pattern may intersect with some obstacles that used to lay inside it, so monotonicity-based methods like binary search cannot be adopted to calculate the genuine $h$. Instead, we first create $C$ with the height equal to the remaining extension requirement and then shrink $h$ until all violations of DRC are eliminated. Multiple DRAs will be separated into independent rouTable~areas and handled independently.

\begin{figure}[b]
    \centering
    \includegraphics[height=1.1in]{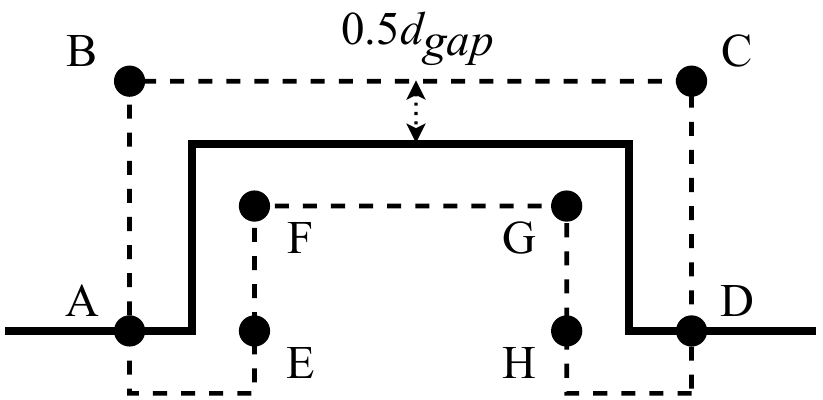}%
    \caption{Illustration of URA. Dash segments represent the URA of a pattern. $ABCD$ is called its outer border, $EFGH$ is called its inner border.}
    \label{ura_example}
\end{figure}

Here, we give the concept of UnReachable Area (URA): The URA of a segment is a rectangle whose border is half of $d_{gap}$ away from the segment, and the URA of a pattern is the union of three segments' URAs, as illustrated in Fig.~\ref{ura_example}. Therefore, we convert DRC into intersection checking between the polygons that stand for URAs or the rouTable~area. Even though the URAs of the previous patterns in the current DP are uncertain, they can be ignored since the validation of state transitions has considered DRC. For convenience, we call edges $AB$ and $CD$ the ``sides'' and $BC$ the ``hat'' of URA. The area below line $AD$ need not be checked because the URA of the original segment certainly lies there, so no other polygons shall exist. During shrinking, we update the height of the outer border $h_{ob}$ of URA and calculate $h$ as:
\begin{equation}
    h = \max\left(0, h_{ob} - \dfrac{d_{gap}}{2}\right)
\end{equation}


The shrinking of $h$ begins with eliminating the violation of DRC for its outer border. In the first place, we shrink $h_{ob}$ according to the intersection of ``sides'' with other polygons. In this paper, we define the distance between a point $p$ and the extended segment $seg$ as $d(seg, p)$, and the distance between a point set $P$ and $seg$ as $d(seg, P) = \min_{p \in P}{d(seg, p)}$. Then the initial shrinking of $h_{ob}$ based on ``side'' is calculated as follows:
\begin{equation}
    h_{ob}^0 = \min\left(d(A, B), d(seg, P_{inters})\right)
    \label{shrink_sides}
\end{equation}
where $P_{inters}$ is the set of all intersection points mentioned above. 

\begin{figure}[t]
    \centering
    \includegraphics[height=1.35in]{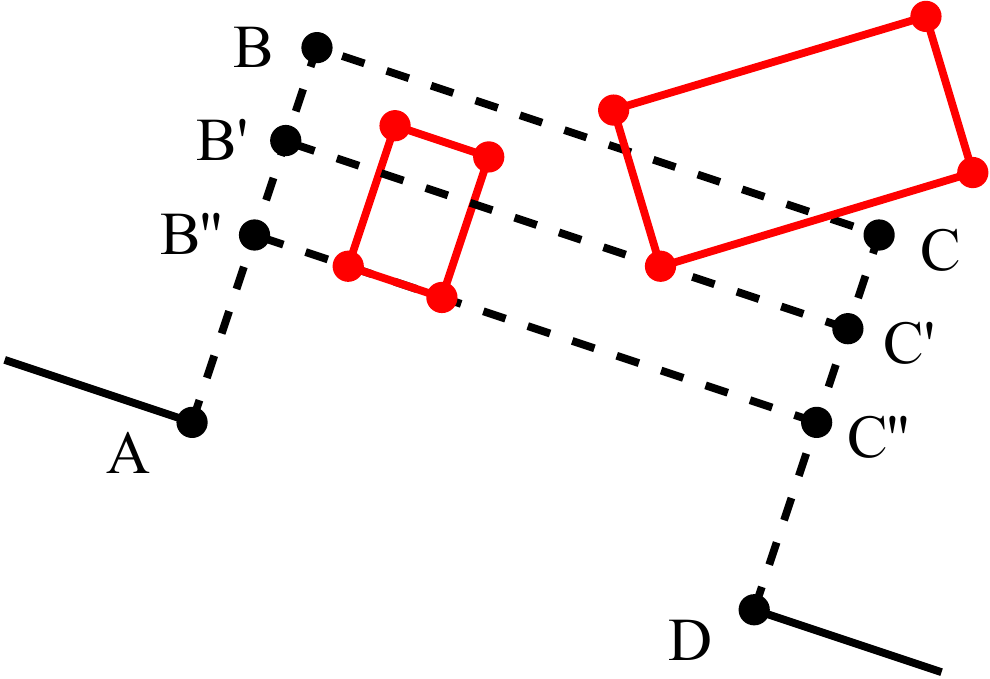}%
    \caption{Illustration of shrinking according to “hat”. For conciseness, we only draw the outer border of URA. $BC$ is shrunk iteratively to $B''C''$ until all intersected polygons are entirely outside the outer border.}
    \label{hat_intersection}
\end{figure}

This way, the violations of DRC caused by the outer border are reduced to the intersection of ``sides'' with other polygons. Shrinking according to ``hat'' may be iterative, as illustrated in Fig.~\ref{hat_intersection}, because the shrunk border may lead to new intersections with other polygons. Observing that there must be at least one node point of each intersected polygon shall remain inside the outer border since the intersection with ``sides'' has been done, we derive an efficient method to solve this problem by checking whether a polygon has node points both inside and outside the outer border, which is presented in Alg.~\ref{hat_checking}. 

\begin{algorithm}[t]
    \SetAlgoLined
    \caption{Shrinking by checking node position.\label{hat_checking}}
    
    \KwIn {$\left\{Poly_{k}\right\}$; Initial URA and $h_{ob}$.}
    \KwOut {Final $P_{inside}$; $h_{ob}$ after shrinking.}
    $P_{check} =\{p\ |\ x_p \in [x_A, x_C], y_p \in [y_D, y_B]\}$ \\
    figure out initial $P_{inside}$ based on $P_{check}$ \\
    \While {$true$}
    {
        \ForEach {$k$}
        {
            update $Poly_{k}^{in}$ \\
            \If {$0 < |Poly_{k}^{in}| < |Poly_{k}|$}
            {
                $h_{ob} = min\left(h_{ob}, d(seg, Poly_{k}^{in})\right)$ \\
                $P_{inside} = P_{inside} \setminus Poly_{k}^{in}$ \\
            }
        }
        \If {$P_{inside} = \bigcup_k{Poly_{k}^{in}}$}
        {
            break
        }
        establish the new outer border using $h_{ob}$ \\
        further reduce $P_{inside}$ according to new outer border \\
    }
\end{algorithm}

We classify all node points of polygon $k$ into set $Poly_{k}$. Defining $P_{inside} = \{p\ |\ p$ is inside the outer border$\}$ and the subset of $Poly_{k}$ named $Poly_{k}^{in} = \{p\ |\ p \in Poly_{k} \cap P_{inside}\}$, then conditions $|Poly_{k}^{in}| = 0$ and $|Poly_{k}^{in}| = |Poly_{k}|$ can indicate the whole polygon $k$ is outside or inside the outer border, respectively, and $h_{ob}$ is updated iteratively as follows: 
\begin{equation}
    \hspace{-0.1in}h_{ob}^{i+1} = \min{\left(h_{ob}^{i}, \mathop{\min}\limits_{k:\ 0 < |Poly_{k}^{in}| < |Poly_{k}|}d(seg, Poly_{k}^{in})\right)}
    \label{shrink_hat}
\end{equation}
We build a segment tree to reduce the checking range from all node points, so that for each URA, we can quickly find a smaller point set $P_{check} = \{p\ |\ x_p \in [x_A, x_C], y_p \in [y_D, y_B]\}$, defining $x_p$ and $y_p$ as the coordinates of $p$, and figure out the initial $P_{inside}$. After each iteration, $P_{inside}$ is reduced to at least $P_{inside} \setminus Poly_{k}$ for all $k$ with $0 < |Poly_{k}^{in}| < |Poly_{k}|$, and the iterating ends when $P_{inside}$ is no longer reduced.

\begin{figure}[t]
    \centering
    \includegraphics[width=3.3in]{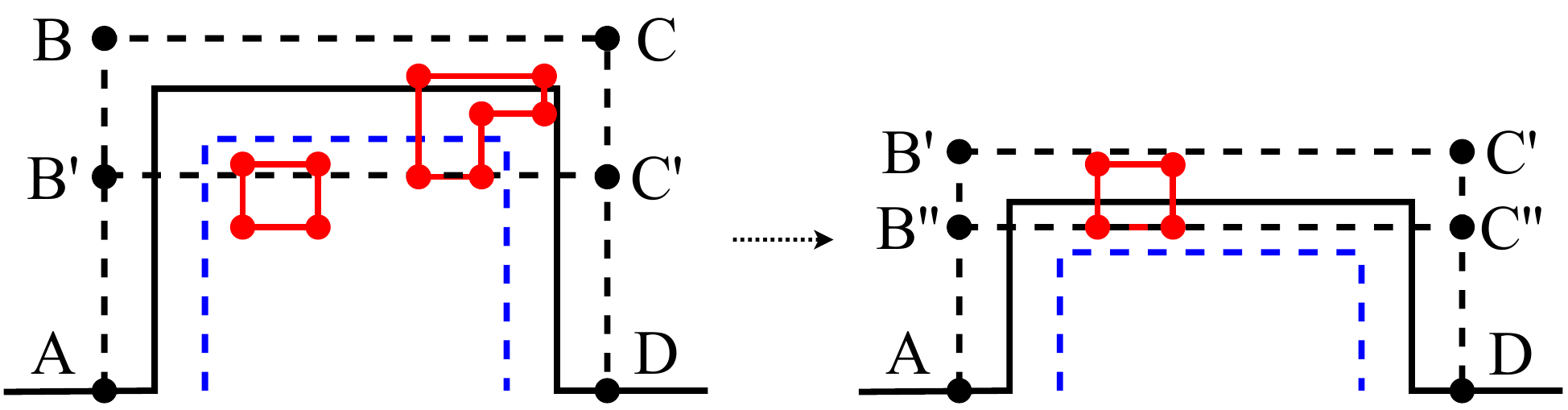}
    \caption{Illustration of the shrinking with the inner border. Black dash segments represent the outer border and its shrinking. Blue dash segments represent the inner border. $BC$ is shrunk iteratively to $B''C''$ until no polygons lay between the inner and outer border.
    }
    \label{inner_intersection}
\end{figure}

For surrounded obstacles, the last work is to check whether polygons inside the outer border intersect the inner border. Similarly, we define the subset of $Poly_{k}$ named $Poly_{k}^{out} = \{p\ |\ p$ is outside the inner border$,\ p \in Poly_{k}\}$. When there exists $Poly_{k}^{out} \cap P_{inside} \neq \emptyset$, $h_{ob}$ must be shrunk to avoid the whole polygon $k$ as follows:
\begin{equation}
    h_{ob}^{i+1} = \min\left(h_{ob}^{i}, \mathop{\min}\limits_{k:\ |Poly_{k}^{out}| > 0}d(seg, Poly_{k})\right)
    \label{shrink_inner}
\end{equation}
This shrinking is also iterative, as shown in Fig.~\ref{inner_intersection}. The difference from the previous one is that $P_{inside}$ is reduced to $P_{inside} \setminus Poly_{k}$ for all $k$ with $|Poly_{k}^{out}| > 0$ here.

Compared to some meandering methods \cite{Yan2009Bsg, Chang2019Obstacle}, our shrinking scheme provides a switchable function to build patterns that route around obstacles if a better state transition is met. Compared to some re-route obstacle-aware methods \cite{Hsu2019dag, Yan2022Single}, our algorithm just slightly changes the topology so that it prevents overriding the original routing.

\subsection{Pattern Restoration}

After the DP, we choose the best state between $dp[n][1]$ and $dp[n][-1]$ as the extended result. Although $dp[i][dir]$ only maintains the value of best extension results, each state has a determined final transition path, so we can easily backtrack it and restore the position of patterns in the best solution.

To this end, we employ a vector corresponding to each state to record the details about how it is obtained at last, which is represented as follows:
\begin{equation}
    transit[i][dir]:\ <i', dir', w'>
\end{equation}
where $i'$ and $dir'$ indicate that the state of $dp[i][dir]$ is transited from the state of $dp[i'][dir']$, and $w'$ denotes that there is a inserted pattern in $dp[i][dir]$ whose feet are located in $u_{i-w'}$ and $u_i$. Additionally, $w' = 0$ marks that this state is not transited through a newly inserted pattern, which is also used to check the extra condition of $p_{local}$ during the state transitions process.

\subsection{Complexity Discussion}

The time complexity of DP state transition is obviously $O(n^2)$, depending on how much the segment is discretized.

The time complexity of pattern restoration is $O(n)$, as we can immediately know the height and width of the pattern to be restored for $dp[i][dir]$ using the information of $transit[i][dir]$, and the length of the transition path is at most $n$.

Suppose $N$ is the total number of node points belonging to the borders of the rouTable~area and the other URAs than which of the current extended segment. As each point is the node of a polygon, its degree must be $2$ so that $N$ also equals the number of segments existing in these polygons. Therefore, the shrinking with ``sides'' is with a time complexity of
\begin{equation}
     O(N) * T(I)
\end{equation}
where $T(I)$ is the time complexity of calculating segment intersection, which can be regarded as $O(1)$ without involving variables.

The time complexity of shrinking with ``hat'' and shrinking with the inner border shall be discussed together.
We employ a segment tree to maintain points whose abscissa rank is within intervals, and the points in each tree node are sorted by ordinate.
The space complexity of this segment tree is $O(Nlog_2{N})$ owing to the fact that each point appears at most $log_2{N}$ times. 
To initialize $P_{check}$, we use abscissa range $[x_A, x_C]$ to conduct a query in the tree and then use binary search to locate the start position of target nodes stored sequential, resulting in a time complexity of $O(4*(\log_2{N}+\log_2{N})) = O(\log_2{N})$.

Suppose $M_r$ and $M_u$ are the numbers of node points in $P_{check}$ belonging to the borders of the rouTable~area and the other URAs, respectively, $M_r$ and $M_u$ will generally be much less than $N$. Define $K$ as the number of sets those $N$ points are classified into. As each iteration will remove at least one of the $K$ sets from $P_{check}$, and all node points belonging to the other URAs are sure to be removed after the first iteration, the following node position checking is with a time complexity of
\begin{equation}
    O(\log_2{N}) + O(KM_r+M_u) * T(R)
\end{equation}
where $T(R)$ is the time complexity of checking whether a point is inside the rectangular inner border. We adopt the ray casting algorithm for this work, whose time complexity can be regarded as $O(1)$ without involving variables.

\section{Multi-Scale Dynamic Time Warping}



\begin{figure}[b]
    \centering
    \includegraphics[width=3.25in]{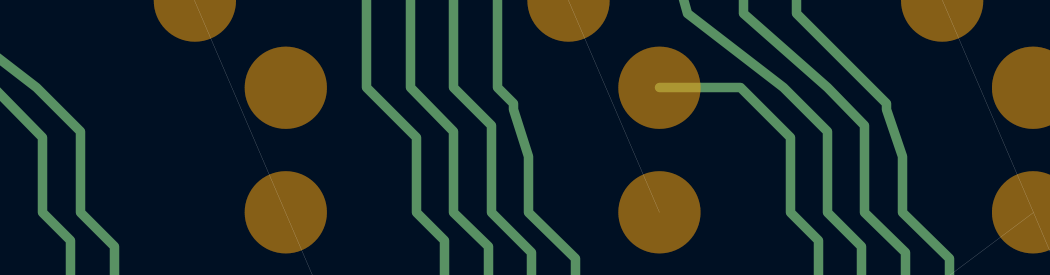}%
    \caption{An example of decoupled differential pairs in the real world.}
    \label{df_not_parallel}
\end{figure}

A matching group may have both single-ended traces and differential pairs. 
During the length matching of such groups, we need not only to match the lengths of all traces, but also to keep the coupling of differential pairs. A common method of length matching involving differential pairs is to regard each differential pair as a wide single-ended trace bounded by its sub-traces. However, this trick still meets many problems in practice because the sub-traces of an actual legal differential pair may frequently not be perfectly coupled in geometry, which may result from not strictly parallel, tiny patterns, or different passed DRAs. Fig.~\ref{df_not_parallel} provides a particular example from a real-world design. 


To tackle these issues, we proposed MSDTW based on the Dynamic Time Warping (DTW) algorithm \cite{Sakoe1978Dynamic, Myers1980Performance}, which converts a differential pair and its DRC into a median single-ended trace. 

\subsection{Considering Nodes instead of Segments}

\begin{figure}[t]
    \centering
    \subfloat[Short segments.]{\includegraphics[width=1.5in]{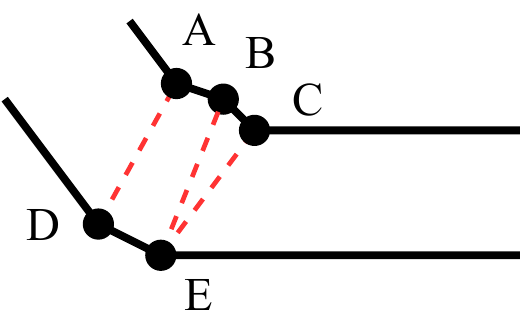}%
        \label{no_parallel_1}}
    \hfil \hfil
    \subfloat[Tiny pattern.]{\includegraphics[width=1.5in]{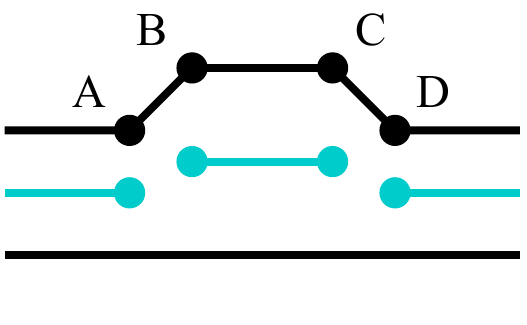}%
        \label{no_parallel_2}}
    \caption{Illustration of coupled sub-traces whose segments are not always strictly parallel. (a) Several short segments appear at a corner. Red dash lines indicate the matching of nodes when DTW is employed. (b) A tiny pattern exists on one of the sub-traces. Cyan lines indicate the expected median segments.}
    \label{no_parallel_issues}
\end{figure}

The conventional method usually uses parallel checking to detect coupled segments of sub-traces so that each pair of coupled segments can be merged into a segment and compose the part of the median trace. This method is effective in theory, but coupled segments of sub-traces may not always be strictly parallel, as illustrated in Fig.~\ref{no_parallel_issues}. Fig.~\ref{no_parallel_1} shows that several short segments appear at a corner because several nodes lie. This case may happen because the coordination of the ideal position can not be represented by the precision of the machine, which makes the ideal nodes be replaced by several approximate nodes. Or it may caused by manual adjustment aiming at avoiding some other objects. Fig.~\ref{no_parallel_2} shows a tiny pattern used for length matching between sub-traces. The pattern causes segments $AB$ and $CD$ not to parallel with the segment of the other sub-trace, and segment $BC$ will further lead the expected median segment shift from its proper position. These two cases are both common in real-world designs.

Although these cases make the coupling imperfect, the differential pair can still be legal in DRC and retained directly. However, these cases bring massive trouble in implementing the parallel checking algorithm, as the implementation must consider a lot of extraordinary case judgments. Even so, it is hard to ensure all possible cases in the future have been covered. From another perspective, we can rely on node matching instead of parallel checking to detect the coupling of segments. Whereas the segments may have some issues in the alignment of angles, the position and clustering of their nodes will not change seriously. Hence, we employ DTW to obtain the optimal node matching between sub-traces except the preserved breakout part. The matching is to minimize the total cost of all matched pairs. The method also allows multiple nodes to match the same node while promising that every node will be matched, which excels at handling the inconsistent number of nodes in two sub-traces, as shown in Fig.~\ref{no_parallel_1}.

Let $trace_P$ and $trace_N$ respectively symbolize the sub-traces in a differential pair. Without loss of generality, we define $C[i][j]$ as the minimum cost of matching the previous $i$ nodes of $trace_P$ and the previous $j$ nodes of $trace_N$. Initializing state $C[0][0] = 0$, the state transition equation of matching is as follows:
\begin{equation}
    C[i][j] = \min\left[
    \begin{array}{lll}
        C[i-1][j] \\
        C[i][j-1] \\
        C[i-1][j-1]
    \end{array}
    \right] + d(i, j),\ \left\{
    \begin{array}{ll}
        i \in [1, I] \\
        j \in [1, J] 
    \end{array}
    \right.
    \label{dtw}
\end{equation}
where $d(i, j)$ is the distance between the $i$th node of $trace_P$ and the $j$th node of $trace_N$, which denotes the matching cost, and $I$ and $J$ denote the number of nodes in the two sub-traces, respectively.

The matched pairs are restored by backtracking the state transition from $C[I][J]$ to $C[0][0]$. We can directly find the current state transits from which of the three predecessor states according to the current matching cost and the minimum cost recorded in these predecessor states. After restoring all matched pairs, we connect every pair of matched nodes, thereby making all nodes compose several connected components. Defining $V_C$ denotes the set of nodes in a connected component, and then $V_C^{P} = \{v\ |\ v \in V_C, v $ is in $ trace_P\}$, $V_C^{N} = \{v\ |\ v \in V_C, v $ is in $ trace_N\}$, we use each $V_C$ to generate a median point $p_m$ as follows:
\begin{equation}
    p_m = \overline{\left\{\overline{V_C^{P}}, \overline{V_C^{N}}\right\}}
    \label{median_point}
\end{equation}
where $\overline{X}$ is the point with the average coordinate of all points in $X$. These median points compose the nodes of the converted median trace. This way, we first respectively calculate the median point of nodes on each sub-trace and then use them to calculate the final median point of a connected component. So that even if multiple nodes are matched to one node, the median points will not shift to one of the sub-traces.

To guarantee the differential pair can be legally restored after length matching, we also attach a virtual DRC to its merged median trace. For a differential pair, the virtual DRC is converted from its distance rule and the original DRC of its sub-traces. Thereby, the restored differential pair will not violate the original DRC as long as the merged median trace does not violate the virtual DRC during length matching.

\subsection{Filtering Unpaired Nodes}

\begin{figure}[t]
    \centering
    \subfloat[Matched pairs involving nodes of tiny patterns.]{\includegraphics[width=3.in]{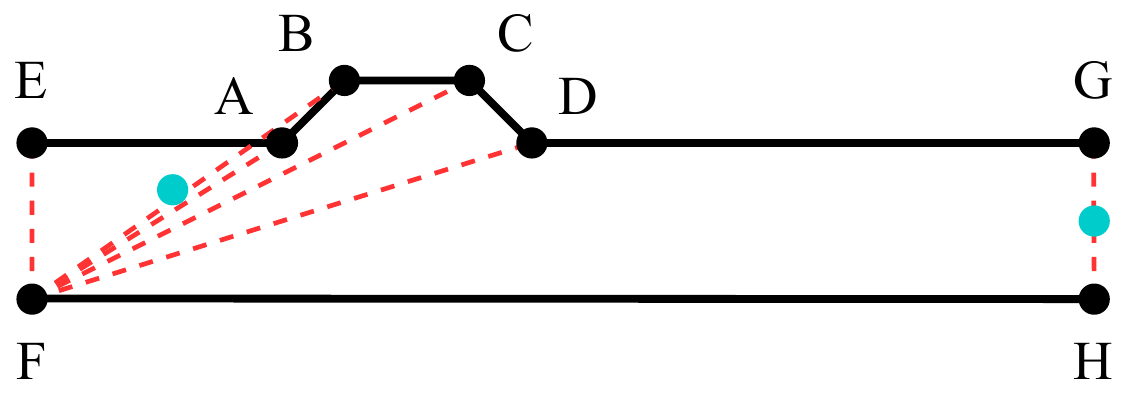}%
        \label{unpaired_nodes_1}}
    \vfil
    \subfloat[Matched pairs after filtering unpaired nodes.]{\includegraphics[width=3.in]{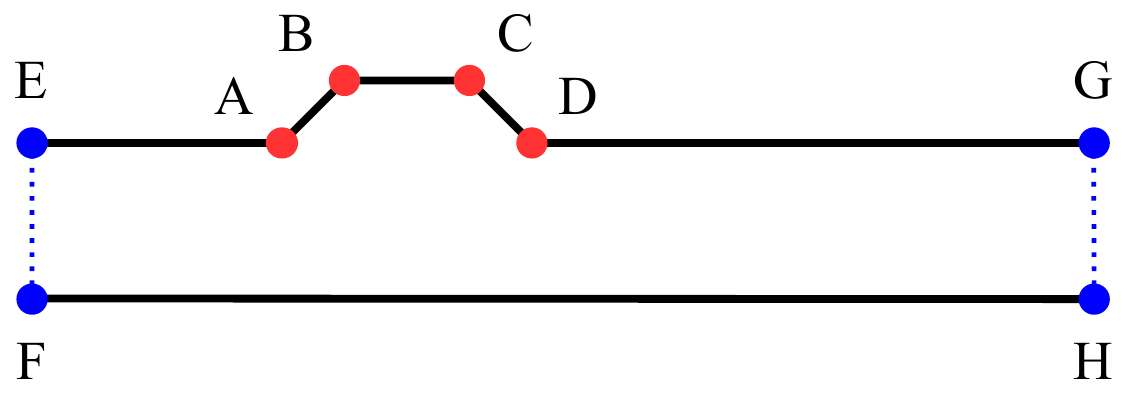}%
        \label{unpaired_nodes_2}}
    \caption{Illustration of the necessity and effect of filtering unpaired nodes. (a) Red dash lines indicate the matching of nodes, and the cyan point indicates the generated median point. (b) Blue dot lines indicate the legally matched pairs. Blue points and red points indicate paired nodes and unpaired nodes, respectively.}
    \label{unpaired_nodes}
\end{figure}
\begin{figure}[t]
    \centering
    \subfloat[If the greatest distance rule is used for matching, the filtering of unpaired nodes may not be controllable.]{\includegraphics[width=2.85in]{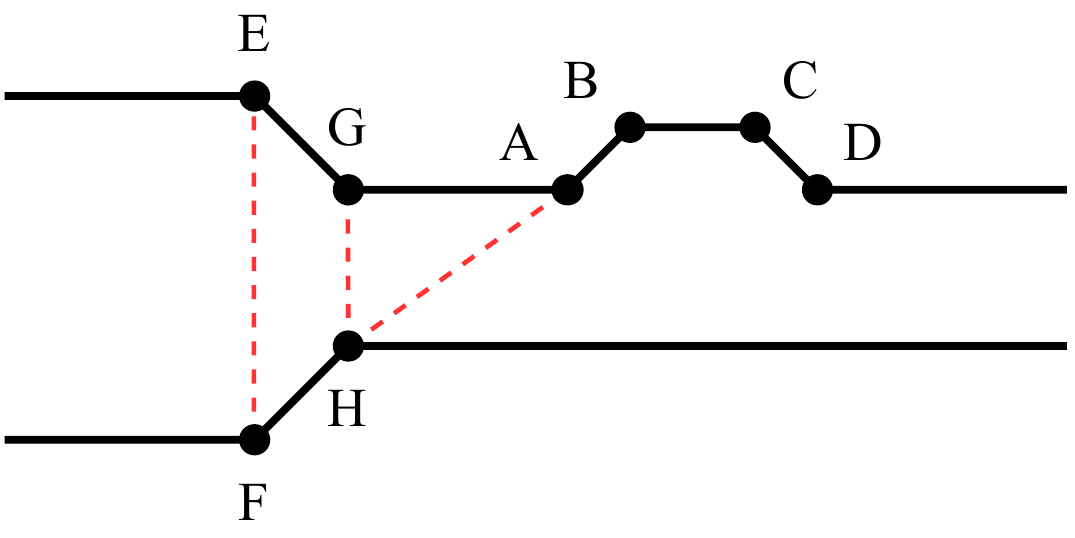}%
        \label{multiple_DRA_1}}
    \hfil \hfil
    \subfloat[MSDTW gradually matches nodes and splits the differential pairs.]{\includegraphics[width=2.85in]{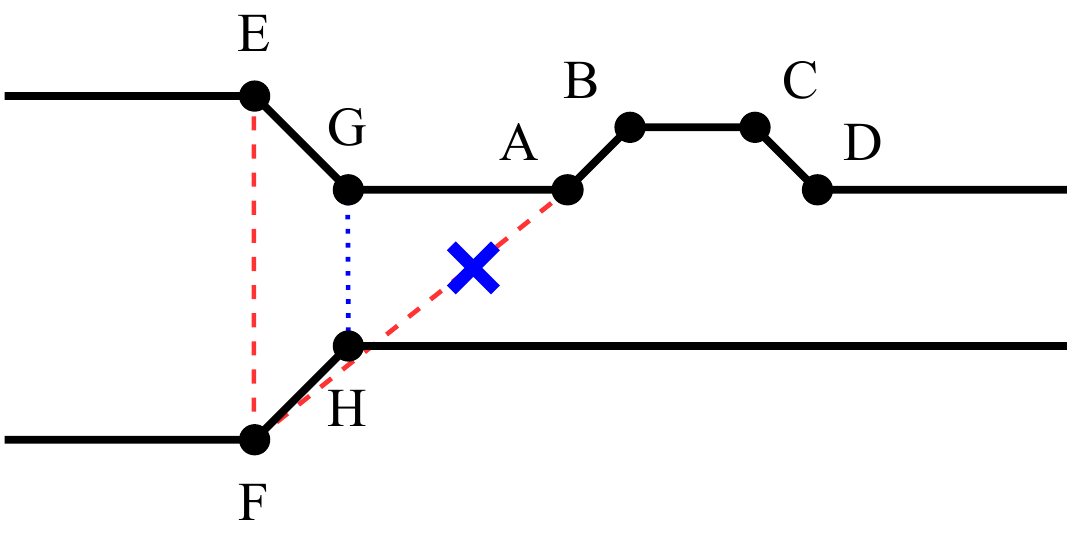}%
        \label{multiple_DRA_2}}
    \caption{Illustration of the issue brought by multiple DRAs and how the MSDTW method tackles it. (a) Red dash lines indicate the possible matched pairs. (b) Blue dot lines indicate the successfully matched pairs in the last round, and red dash lines indicate the possible matched pairs in this round.}
    \label{multiple_DRA}
\end{figure}

Matched pairs involving nodes of tiny patterns usually cause an undesirable shifting of median points, as illustrated in Fig.~\ref{unpaired_nodes_1}. The naive DTW will find a matched pair for all nodes, even including those of tiny patterns. When matched pairs involve nodes of tiny patterns, the corresponding connected component will generate a seriously shifted median point according to Eq.~\eqref{median_point}. Hence, the nodes of tiny patterns shall be regarded as noise during the running of DTW.

To avoid them, defining $cost_{i}$ as the matching cost of the matched pair $pair_{i}$ and $r$ as the distance rule of the differential pair, we will drop $pair_{i}$ if $cost_{i} > \sqrt{2}r$. Considering the rotation angle of a trace must be obtuse, any matched pair, even if at a corner, shall meet $cost_{i} > \sqrt{2}r$, otherwise we can determine that it is a matched pair involving nodes of tiny patterns. The nodes that only belong to the dropped matched pairs are called unpaired nodes, and the remaining nodes are called paired nodes. Consequently, all unpaired nodes will be filtered and no longer influence the position of generated median points, as shown in Fig.~\ref{unpaired_nodes_2}.

\subsection{Tackling Multiple Design Rule Areas}

\begin{algorithm}[t]
    \SetAlgoLined
    \caption{Multi-Scale Dynamic Time Warping.\label{msdtw}}
    
    \KwIn {Differential pair $df_{original}$; Rule set $R$.}
    \KwOut {Set of all matched pairs $M$.}
    $M = \emptyset$, set of differential sub-pairs $S = \{df_{original}\}$ \\
    \ForEach {$r$ in $R$}
    {
        \ForEach {$df_{i}$ in $S$}
        {
            calculate the current matched pairs set $M_{i}$ \\
            \ForEach {$pair_{j}$ in $M_{i}$}
            {
                \If {$cost_{j} > \sqrt{2}r$}
                {
                    $M_{i} = M_{i} \setminus \{pair_{j}\}$ \\
                }
            }
            $M = M \cup M_{i}$ \\
            split $df_{i}$ into $S_{split}$ using matched pairs in $M_{i}$ \\
            \ForEach {$df_{split}$ in $S_{split}$}
            {
                \If {no nodes in $trace_P$ or $trace_N$}
                {
                    $S_{split} = S_{split} \setminus \{df_{split}\}$ \\
                }
            }
            $S = S_{split} \cup S  \setminus \{df_{i}\}$ \\
        }
    }
\end{algorithm}

\begin{figure}[b]
    \centering
    \includegraphics[height=0.95in]{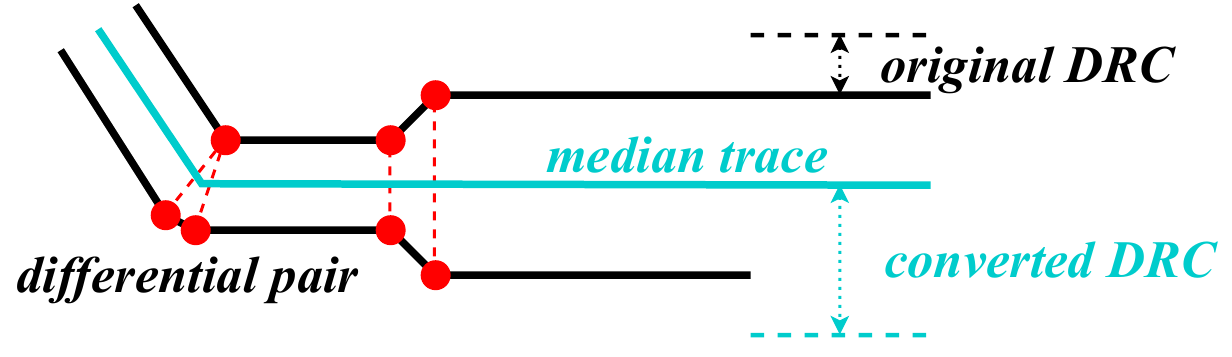}
    \caption{Final median trace of a differential pair merged by MSDTW. Red dash lines in the figure indicate matched pairs. 
    }
    \label{dtw_match}
\end{figure}

The filtering of unpaired nodes relies on the distance rule of differential pairs, but if a differential pair passes multiple DRAs, the corresponding multiple distance rules may cause filtering failure. As illustrated in Fig.~\ref{multiple_DRA_1}, node pair $EF$ and $GH$ belong to different DRAs. If we use $d(E, F)$ as the distance rule for matching, the unpaired node $A$ cannot be filtered since $d(A, H) < \sqrt{2} * d(E, F)$. And if we use $d(A, H)$ as the distance rule, the correctness of matching in other DRAs cannot be promised. 

To handle differential pairs passing multiple DRAs, our MSDTW method sequentially matches the nodes by increasing distance rules, the so-called multi-scale. Defining $R = \{r_0, r_1, ..., r_m\}$ as the set of all involved distance rules whose elements are arranged in increasing order, we match the nodes by enumerating the elements of $R$ and continuously filter unpaired nodes after each round of matching. 

In the beginning, we match nodes by the smallest $r_0 \in R$, so that we can temporarily prevent the matching from involving any node of tiny patterns. Once the matching of the current round is determined, we split the remaining differential pair into differential sub-pairs according to the matched nodes in this round. If there has been no node on $trace_P$ or $trace_N$ of a sub-pair, this sub-pair will be dropped immediately since no more meaningful matching can occur. This dropping strategy holds because tiny patterns are used for length matching between sub-traces and shall only appear on either $trace_P$ or $trace_N$, or else two tiny patterns in different sub-traces can be reduced by each other. 

This way, we can filter all unpaired nodes in the DRA corresponding to $r_0$, thus isolating them from the matching of the subsequent rounds. In the next round, we select the second smallest $r_1 \in R$ and match nodes individually in each retained sub-pair, which means that the matching of nodes across sub-pairs is also forbidden. As shown in Fig.~\ref{multiple_DRA_2}, by a smaller distance rule, node pair $GH$ is successfully matched while node pair $AH$ is dropped. Then, $GH$ splits the original differential pair into two sub-pairs so that the matching of node pair $AF$ is also forbidden.

This recursion ends when no sub-pair is further split, and then we collect all successfully matched pairs as the result. At this point, we can present the whole MSDTW method in Alg.~\ref{msdtw} and its illustration in Fig.~\ref{dtw_match}. After length matching, we restore the differential pairs and compensate tiny patterns to sub-traces if needed.



\section{Experiments}

Our length-matching tool is developed using C++ programming language. All experiments were performed with an AMD Ryzen 7840H 3.80 GHz CPU and 16GB memory. 

\subsection{Overall Length-Matching Performance}

\renewcommand{\arraystretch}{1.35}

\begin{table*}[t]
    \centering
    \caption{Length-matching performance Compared with Allegro \cite{APD} AiDT}
    \label{experiment}
    \begin{tabular}{|c|c|c|c|c|c|c|c|c|c|c|c|c|}
        \hline
        \multirow{2}{*}{\makecell{\vspace{-0.11in} \\ case}} & \multirow{2}{*}{\makecell{\vspace{-0.135in} \\ $\dfrac{l_{target}}{d_{gap}}$}} & \multirow{2}{*}{\makecell{\vspace{-0.11in} \\ group size}} & \multirow{2}{*}{\makecell{\vspace{-0.11in} \\ trace type}} & \multirow{2}{*}{\makecell{\vspace{-0.11in} \\ spacing}} & \multicolumn{3}{c|}{Max. error (\%)} & \multicolumn{3}{c|}{Avg. error (\%)} & \multicolumn{2}{c|}{runtime (s)} \\
        \cline{6-13}
        & & & & & \multicolumn{1}{c|}{Initial} & \multicolumn{1}{c|}{Allegro} & \multicolumn{1}{c|}{Ours} & \multicolumn{1}{c|}{Initial} & \multicolumn{1}{c|}{Allegro} & Ours & \multicolumn{1}{c|}{Allegro} & Ours \\
        \hline
        1 & 205.88 & 8 & single-ended & dense  & \multicolumn{1}{c|}{37.38} & \multicolumn{1}{c|}{33.52}          & \multicolumn{1}{c|}{\textbf{3.02}} & \multicolumn{1}{c|}{19.02} & \multicolumn{1}{c|}{14.23} & \textbf{1.30} & \multicolumn{1}{c|}{\textbf{0.92}} & \multicolumn{1}{c|}{6.87}          \\ 
        \hline
        2 & 199.02 & 8 & single-ended & dense  & \multicolumn{1}{c|}{35.99} & \multicolumn{1}{c|}{28.06}          & \multicolumn{1}{c|}{\textbf{3.93}} & \multicolumn{1}{c|}{19.41} & \multicolumn{1}{c|}{11.04} & \textbf{1.39} & \multicolumn{1}{c|}{\textbf{0.78}} & \multicolumn{1}{c|}{3.98}          \\ 
        \hline
        3 & 187.25 & 8 & single-ended & dense  & \multicolumn{1}{c|}{35.91} & \multicolumn{1}{c|}{20.91}          & \multicolumn{1}{c|}{\textbf{3.51}} & \multicolumn{1}{c|}{20.06} & \multicolumn{1}{c|}{8.66}  & \textbf{1.37} & \multicolumn{1}{c|}{\textbf{0.81}} & \multicolumn{1}{c|}{5.27}          \\ 
        \hline
        4 & 186.27 & 8 & single-ended & dense  & \multicolumn{1}{c|}{30.99} & \multicolumn{1}{c|}{22.25}          & \multicolumn{1}{c|}{\textbf{5.46}} & \multicolumn{1}{c|}{17.22} & \multicolumn{1}{c|}{9.85}  & \textbf{1.83} & \multicolumn{1}{c|}{\textbf{0.72}} & \multicolumn{1}{c|}{2.86}          \\ 
        \hline
        5 & 217.32 & 4 & differential & sparse & \multicolumn{1}{c|}{26.55} & \multicolumn{1}{c|}{\textbf{10.21}} & \multicolumn{1}{c|}{10.3}          & \multicolumn{1}{c|}{15.18} & \multicolumn{1}{c|}{5.14}  & \textbf{3.32} & \multicolumn{1}{c|}{5.07}          & \multicolumn{1}{c|}{\textbf{3.22}} \\ 
        \hline
    \end{tabular}
\end{table*}

The benchmark in this subsection is derived from the sample design provided by Allegro PCB Designer \cite{APD}, in which we removed the tuning of preset matching groups and arranged 5 cases to evaluate the overall performance of length matching. Since we knew no published works concerned with the same objective and constraints of this problem, we compared our approach with the SOTA Auto-interactive Delay Tune (AiDT) function of Allegro PCB Designer, which is also applied in real-world industrial designs against the problem in this paper. The comparison is measured by the matching error of each matching group and the runtime of the algorithm, the metrics of matching error are detailed as follows:
\begin{equation}
    \begin{array}{lll}
        \text{Max. error} & = & \mathop{\max}\limits_{i \in [1, L]}\dfrac{{l_{target} - l_{i}}}{l_{target}} \\
        \vspace{-0.1in} \\
        \text{Ave. error} & = & \dfrac{\sum_i^L (l_{target} - l_{i})}{n \times l_{target}} \\
    \end{array}
\end{equation}
where $l_{i}$ is the length of the $i$th trace after length matching and $L$ is the number of traces in the case.

\begin{figure}[t]
    \centering
    \subfloat[Our experimental result.]{\includegraphics[width=3.15in]{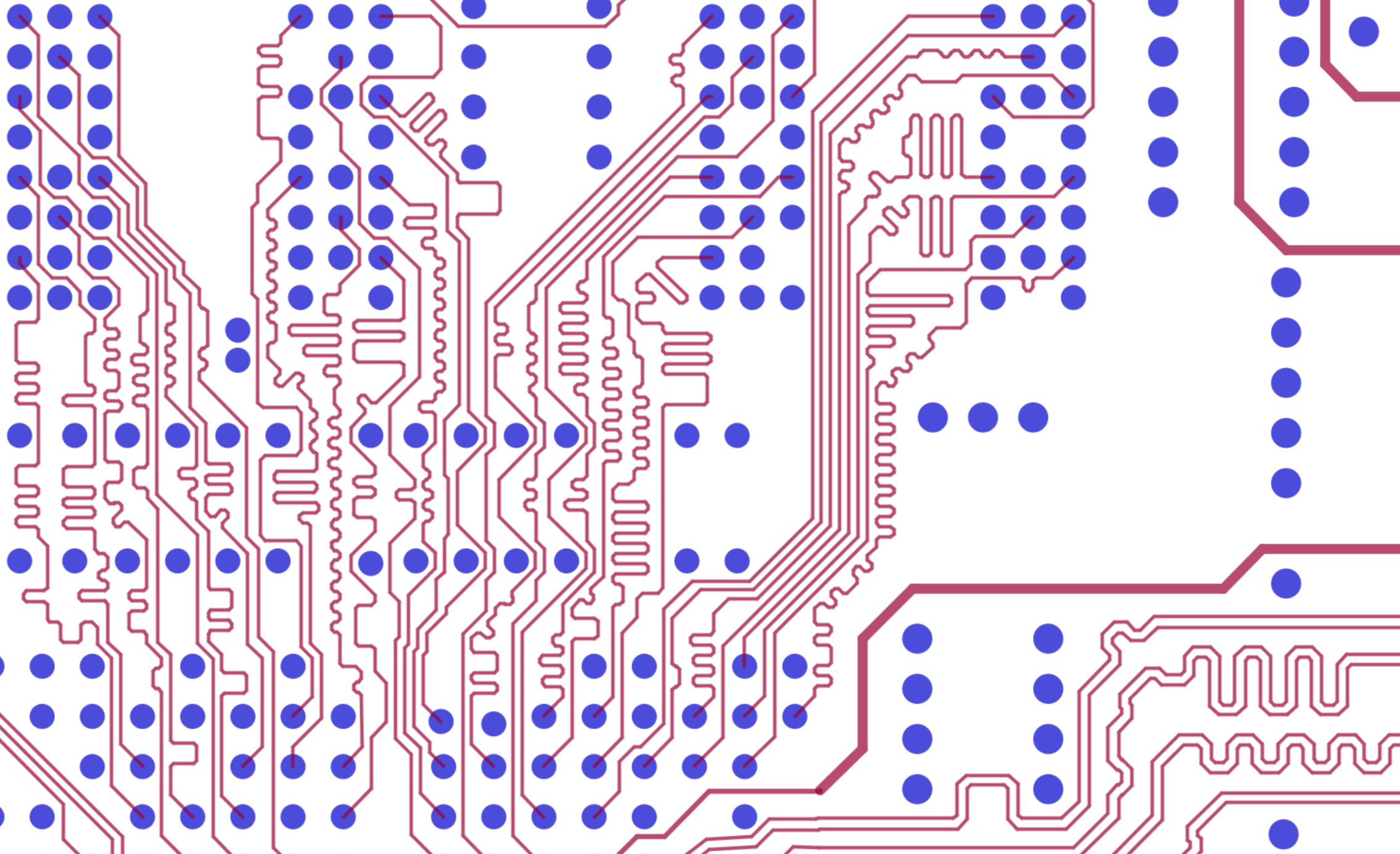}%
        \label{cadence_demo}}
    \vfil
    \subfloat[Any direction functionality.]{\includegraphics[width=3.15in]{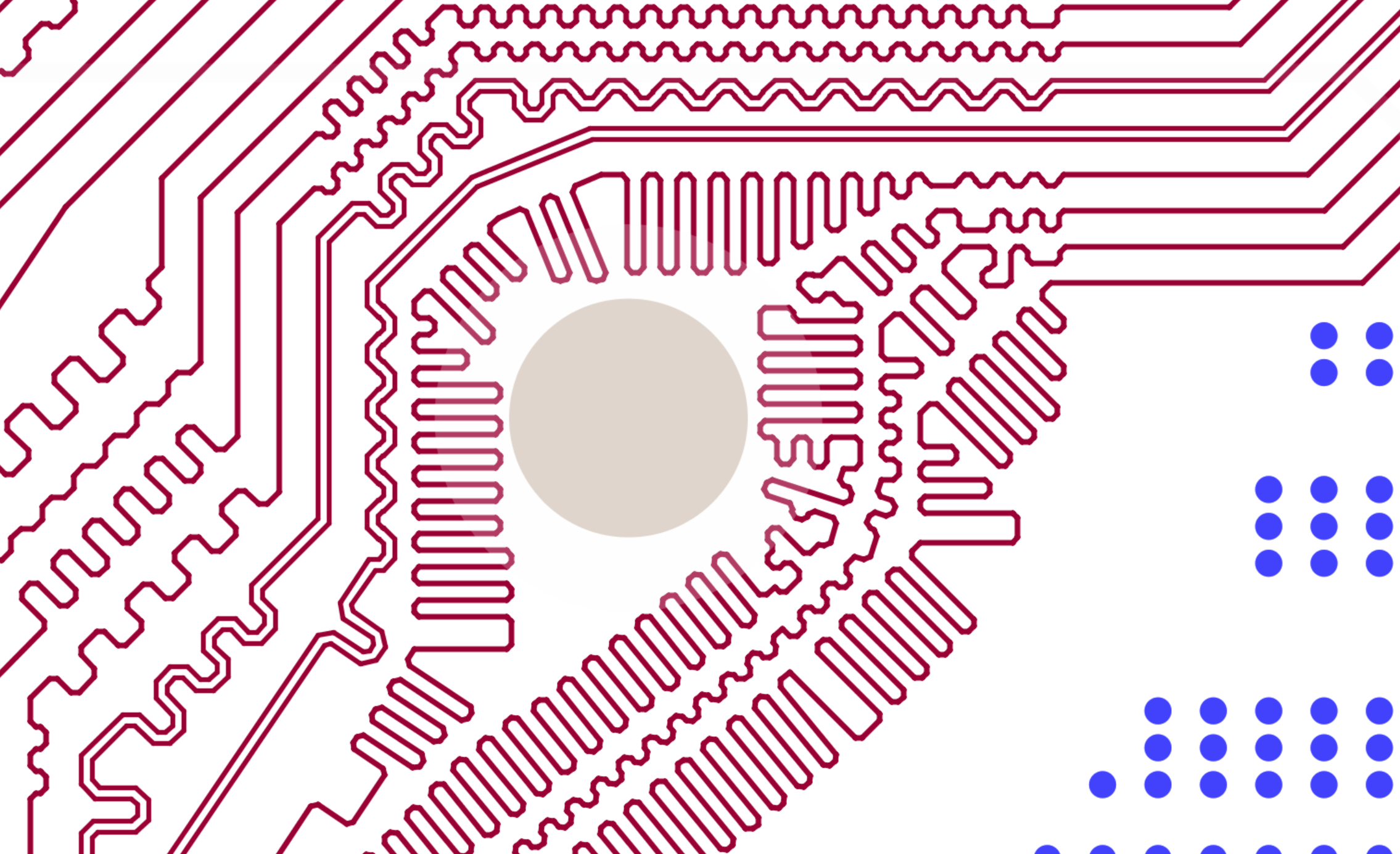}%
        \label{any_trace}}
    \caption{Displays of our length-matching results.}
    \label{experiment_figure}
\end{figure}

Table~\ref{experiment} gives case statistics and experimental results. As shown in the table, our approach addresses more precise length matching in the first 4 cases while compromising on runtime, which resulted from our DP-based extension having more flexible space utilization in spacing-dense environments and inevitable time complexity. Nevertheless, the presented runtime is still reasonable for human tolerance, and on the contrary, manually refining the delay tuning is usually pretty time-consuming. In case 5, our approach shows a similar length-matching precision to AiDT when spacing is sparse but wins an advantage in runtime. As the AiDT algorithm of Allegro is not public, we can only infer that it owes to the better efficiency of our MSDTW method. 

Fig.~\ref{cadence_demo} shows our experimental result, and Fig.~\ref{any_trace} gives a dummy routing on a modified private commercial design that shows our functionality with any-direction traces.

\subsection{Ablation Experiments of DP}

\begin{table}[t]
    \centering
    \caption{Extension performance with and without DP}
    \label{ablation_DP}
    \begin{tabular}{|c|c|c|c|c|}
        \hline
        \multirow{2}{*}{\makecell{\vspace{-0.11in} \\ case}} & \multirow{2}{*}{\makecell{\vspace{-0.135in} \\ $\dfrac{d_{gap}}{w_{trace}}$}} & \multirow{2}{*}{\makecell{\vspace{-0.135in} \\ $\dfrac{l_{original}}{d_{gap}}$}} & \multicolumn{2}{c|}{extension upper bound (\%)} \\
        \cline{4-5}
        & & & \multicolumn{1}{c|}{with DP} & \multicolumn{1}{c|}{without DP} \\
        \hline
        1 & 2.5 & 24.89 & \multicolumn{1}{c|}{879.30} & \multicolumn{1}{c|}{845.80} \\  
        \hline
        2 & 3.0 & 21.33 & \multicolumn{1}{c|}{718.79} & \multicolumn{1}{c|}{742.16} \\  
        \hline
        3 & 3.5 & 18.67 & \multicolumn{1}{c|}{581.42} & \multicolumn{1}{c|}{345.62} \\  
        \hline
        4 & 4.0 & 16.59 & \multicolumn{1}{c|}{481.14} & \multicolumn{1}{c|}{229.79} \\  
        \hline
        5 & 4.5 & 14.93 & \multicolumn{1}{c|}{428.33} & \multicolumn{1}{c|}{177.92} \\  
        \hline
        6 & 5.0 & 13.57 & \multicolumn{1}{c|}{327.41} & \multicolumn{1}{c|}{ 80.20} \\  
        \hline
    \end{tabular}
\end{table}

\begin{figure*}[t]
    \centering
    \subfloat[Case 1 with DP.]{\includegraphics[width=2.15in, angle=0]{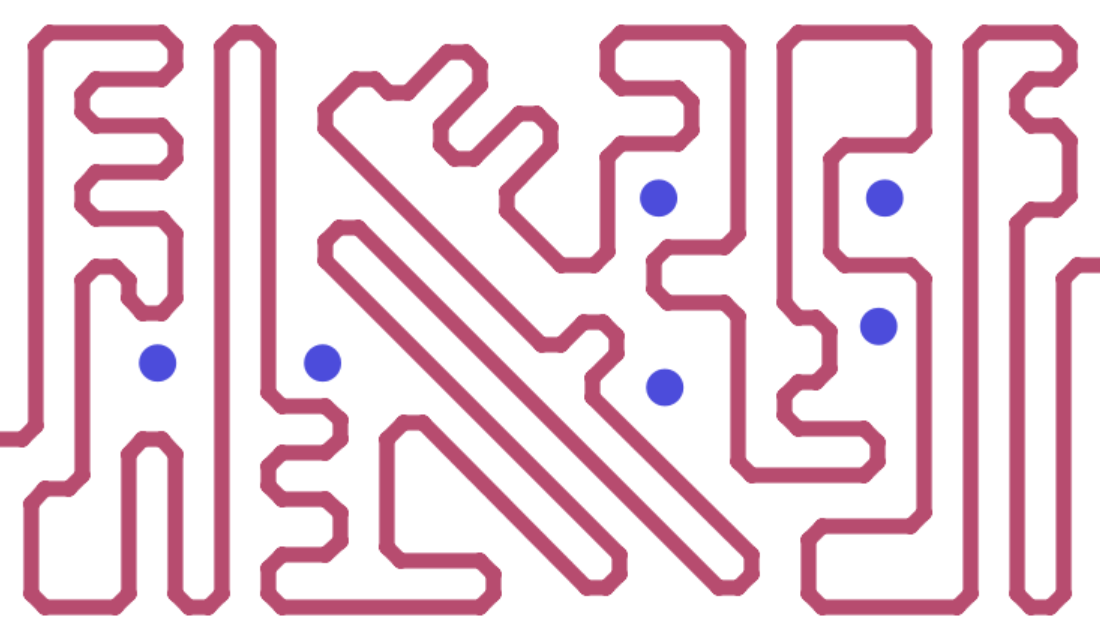}%
        \label{DP1}}
    \hfil \hspace{0.1in}
    \subfloat[Case 5 with DP.]{\includegraphics[width=2.15in, angle=0]{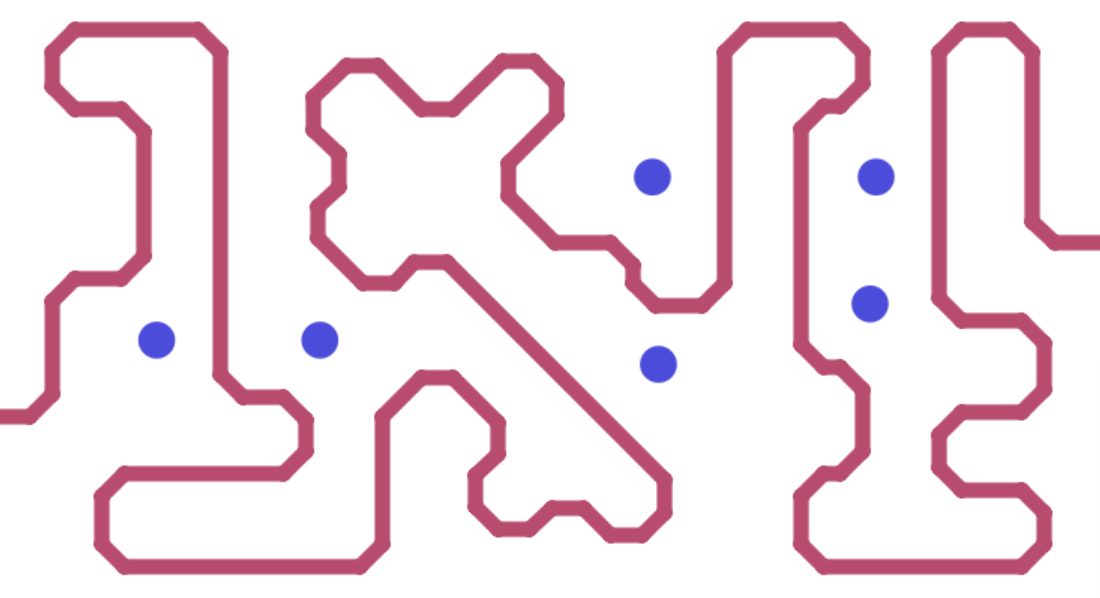}%
        \label{DP5}}
    \hfil \hspace{0.1in}
    \subfloat[Case 6 with DP.]{\includegraphics[width=2.15in, angle=0]{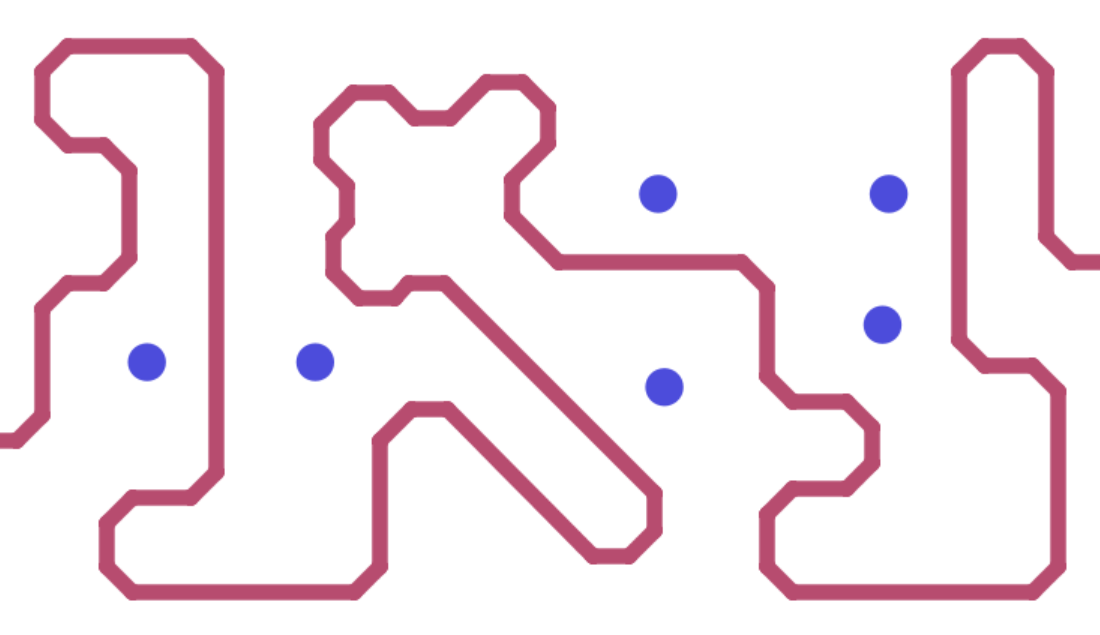}%
        \label{DP6}}
    \vfil
    \subfloat[Case 1 without DP.]{\includegraphics[width=2.15in, angle=0]{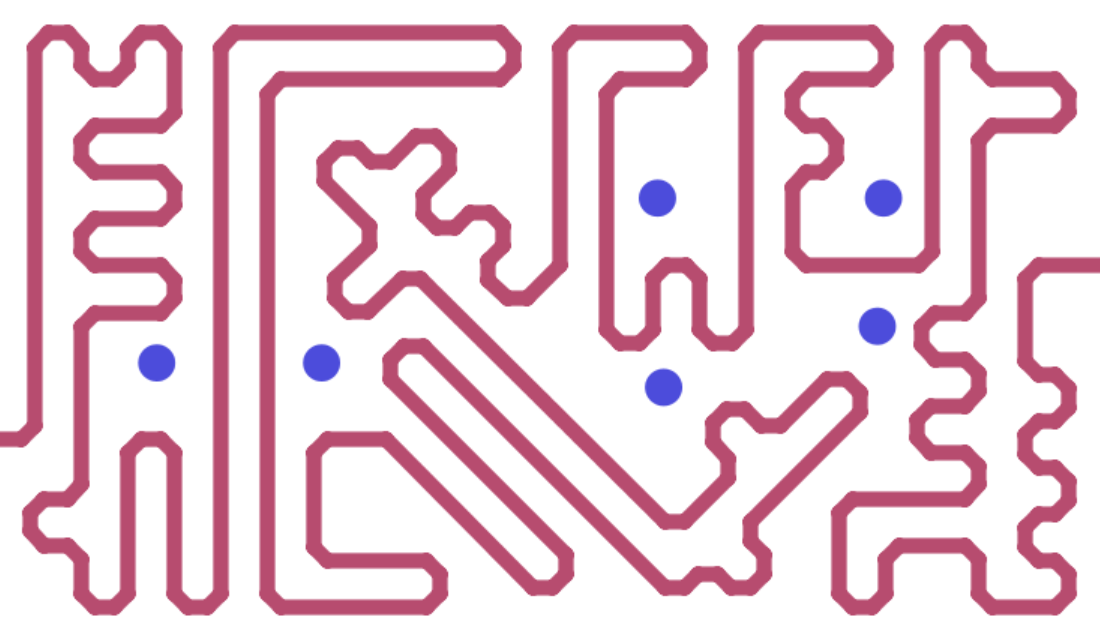}%
        \label{abl1}}
    \hfil \hspace{0.1in}
    \subfloat[Case 5 without DP.]{\includegraphics[width=2.15in, angle=0]{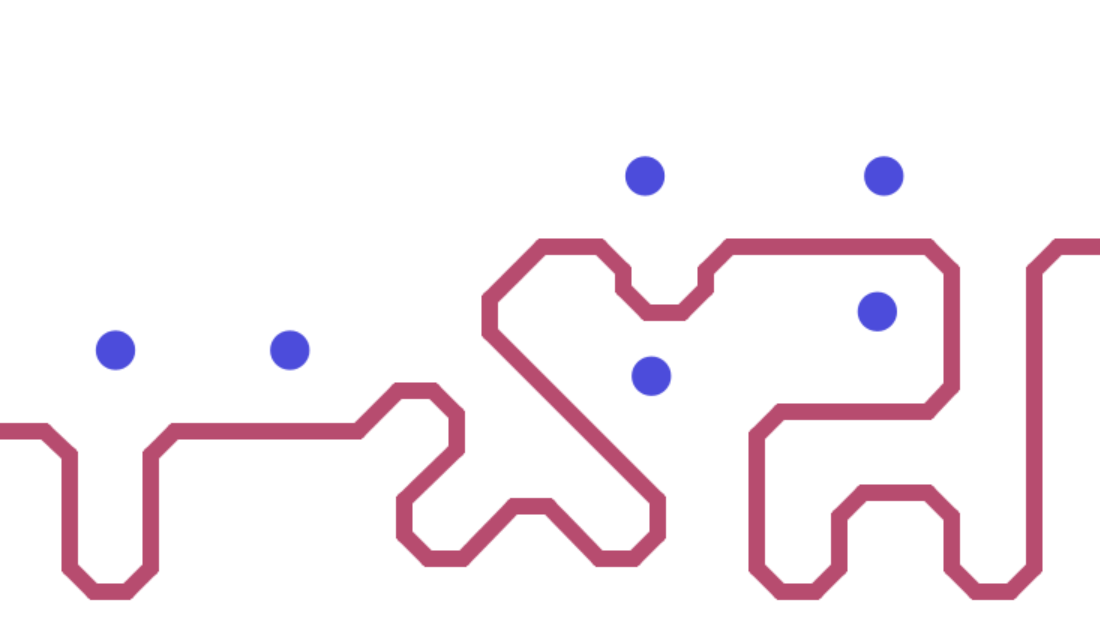}%
        \label{abl5}}
    \hfil \hspace{0.1in}
    \subfloat[Case 6 without DP.]{\includegraphics[width=2.15in, angle=0]{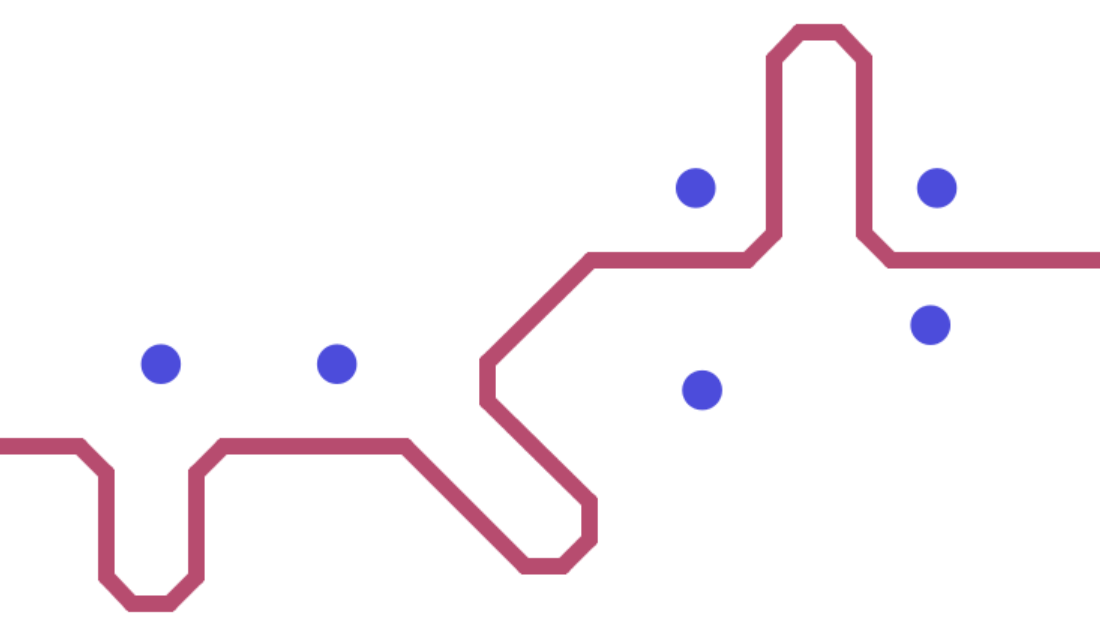}%
        \label{abl6}}
    \caption{Displays of extension performance with and without DP.}
    \label{ablation_DP_figure}
\end{figure*}

To examine the obstacle awareness of the proposed DP algorithm, we further conduct ablation experiments on a dummy design with narrow space between dense vias. The compared algorithm without DP is based on fixed routing tracks and constant pattern width. The statistics and results of experimental cases are given in Table~\ref{ablation_DP}. In this table, $w_{trace}$ denotes the width of the extended trace, and $l_{original}$ denotes the trace length before length matching. The second column shows that we gradually increase $d_{gap}$ to strengthen the restriction of DRC during the experiment. The third column indicates the ideal number of patterns that can be directly inserted perpendicular to the original according to the ratio of $l_{original}$ to $d_{gap}$. We evaluate the performance by measuring the upper bound of the extended length compared to the original length, which is calculated as:
\begin{equation}
    \dfrac{l_{extended} - l_{original}}{l_{original}} \times 100\%
\end{equation}
where $l_{extended}$ is the trace length after length matching. Typically, we display three results of the cases in Fig.~\ref{ablation_DP_figure} to incorporate with the numerical metric.

It can be observed that the performance of the two algorithms is quite similar before the DRC restriction is strengthened enough. The algorithm with DP performs 3.96\% better than the one without DP in case 1 while 3.15\% worse in case 2. It is noteworthy that the result of case 2 is also reasonable because the former algorithm is essentially DP plus greedy, and the DP part is adopted for achieving the local optimum of segments but does not promise the global optimum of the whole trace.

In case 5, the algorithm without DP fails to utilize the space above the trace, and it also does not sufficiently utilize the space in the lower left area. It results from the algorithm cannot flexibly choose the patterns' feet or adjust the patterns' width. In this case, most of the fixed tracks leading to the upper area happen to be too close to the obstacles. The right part of the 135-degree segment in the middle seems can hold a pattern extended to the upper area, but the right foot of this pattern will actually violate $d_{protect}$ away from the node of the original segments. Meanwhile, the empty space in the lower left part is just smaller than the square of $2d_{gap} \times 2d_{gap}$, so it is unable to further contain any pattern. While turning to the DP algorithm, these issues are all resolved properly. In the left and right areas, the algorithm wisely adjusts the patterns' width, routes around obstacles, and connects two patterns in opposite directions, which produces more space in the lower left part to contain a couple of continuous patterns. As for the middle part, the algorithm chooses the node of the original segments as a foot of the pattern, thereby avoiding the trouble of $d_{protect}$. All these improvements led by DP result in a significant advantage compared to another algorithm.

\subsection{MSDTW} 

\begin{figure}[t]
    \centering
    \subfloat[An original differential pair (white) and its merged median trace (green).]
    {\includegraphics[width=2.5in, angle=270, clip, trim=0 0 1.5in 0]{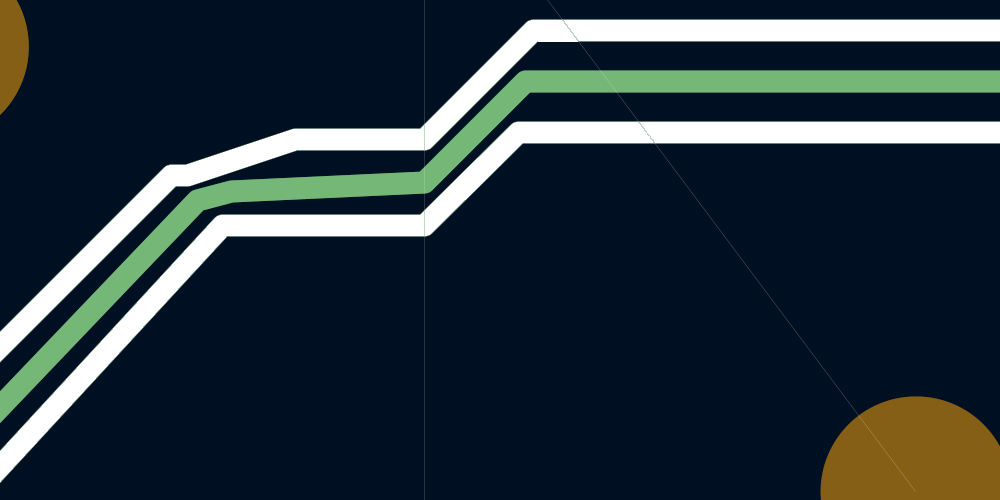}%
        \label{df_merge}}
    \hfil \hspace{0.15in}
    \subfloat[A median trace (white) and its restored differential pair (green).]
    {\includegraphics[width=2.5in, angle=270, clip, trim=0 0 1.5in 0]{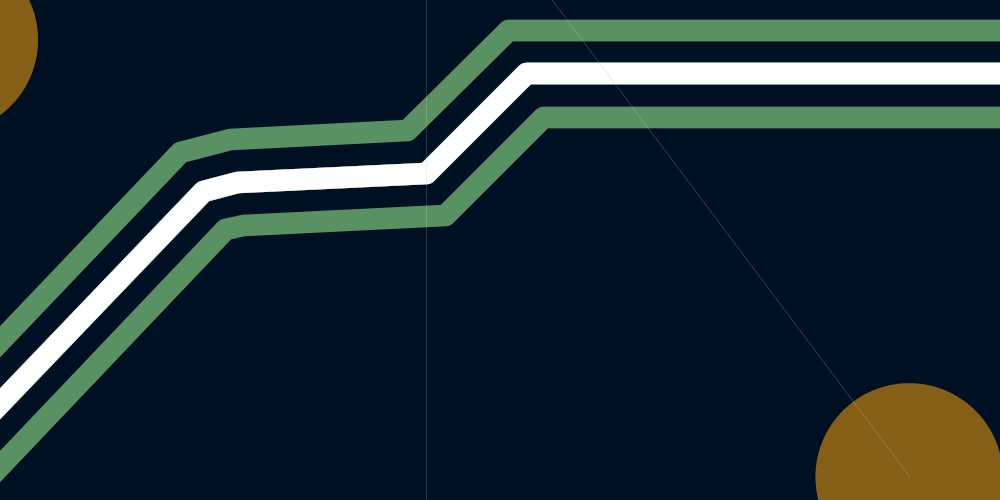}%
        \label{df_restore}}
    \caption{Example of the functionality of MSDTW based on the case from the design in Fig.~\ref{df_not_parallel}.}
    \label{df}
\end{figure}

Fig.~\ref{df_merge} and Fig.~\ref{df_restore} display the merged median trace and restored differential pair of a case from the design in Fig.~\ref{df_not_parallel}, respectively.

\section{Conclusion}

In this paper, we present an automatic length-matching approach concerning any-direction traces in high-speed designs. Unlike previous works, we employ DP and computational geometry to meander the trace, which aims to preserve the specified original routing during length matching by maintaining direction and limiting overriding changes in topology. Meanwhile, it achieves more flexible obstacle-aware space utilization with reasonable runtime. Moreover, we proposed a method named MSDTW that converts differential pairs during length matching to tackle the issues against decoupling and multiple DRAs. The experimental illustration and the comparison with commercial tools demonstrate the effectiveness and functionality of our approach.

\section*{Acknowledgments}

This work is supported by the National Natural Science Foundation of China (No. 12271098).


 





\end{document}